\documentclass[useAMS,usenatbib]{mn2e}
\usepackage {graphicx}

\title[A kinematic study of open clusters: implications for their origin]{A kinematic study of open clusters: implications for their origin}

\author[D. Vande Putte, T.P. Garnier, I. Ferreras, R.P. Mignani, and Mark Cropper]{D. Vande Putte,\thanks{E-mail:
dwvp@mssl.ucl.ac.uk (DVP)} T.P. Garnier,  I. Ferreras, R.P. Mignani, and Mark Cropper\\  
University College London, Mullard Space Science Laboratory, Holmbury St Mary, Dorking, RH5 6NT, UK\\}

\begin{document}

\date{Accepted yyyy mm dd. Received  yyyy mm dd; in original form  yyyy mm dd}

\pagerange{\pageref{firstpage}--\pageref{lastpage}} \pubyear{2002}

\maketitle

\label{firstpage}

\begin{abstract}

The Galactic population of open clusters provides an insight into star formation in the Galaxy.  The open cluster catalogue by Dias et al.(2002b) is a rich source of data, including kinematic information.  This large sample made it possible to carry out a systematic analysis of 481 open cluster orbits, using parameters based on orbit eccentricity and separation from the Galactic plane.   These two parameters may be indicative of origin, and we find them to be correlated.  We also find them to be correlated with metallicity, another parameter suggested elsewhere to be a marker for origin in that high values of any of these two parameters generally indicates a low metallicity ([Fe/H] Solar$<-$0.2 dex).  The resulting analysis points to four open clusters in the catalogue being of extra-Galactic origin by impact of high velocity cloud on the disk: Berkeley21, 32, 99, and Melotte66, with a possible further four due to this origin (NGC2158, 2420, 7789, IC1311).  A further three may be due to Galactic globular cluster impact on the disk i.e of internal Galactic origin (NGC6791, 1817, and 7044).  

\end{abstract}


\section{Introduction}

In general, Galactic open clusters are ensembles of stars with low central
concentration and irregular shape. They constitute the outcome of a
star forming region and are found across a range of evolving
stages. Open clusters arguably represent the growth of the disk of the
Galaxy via a smooth, self-regulated process of star formation, in
contrast with globular clusters -- formed during the early stages of
the Milky Way -- or with young massive clusters, found in starbursting
galaxies. The fundamental difference between open and globular
clusters appears to be the pressure of the environment, with open
clusters formed in regions with low pressures (Elmegreen \& Efremov
1997). The complex formation history of the Milky Way, characterized
by a number of events with enhanced star formation along with a smooth
low-level constant formation (de la Fuente Marcos \& de la Fuente
Marcos 2004) should be imprinted in the ``fossil record" represented by
the open clusters, the globular clusters and the bulk of the stellar
population in the Galaxy. 

The total mass of an open cluster ranges
from a few tens to a few thousands of solar masses (see for example Durgapal
\& Pandey 2001), and have typical diameters of a few to a few tens of
pc.  Their stars form as dense clusters located within the parental
giant molecular cloud (Pfalzner 2009).  Unlike the case of globular
clusters, the population of open clusters is very strongly
concentrated along the Galactic plane, which also makes them difficult
to observe at large distances. As they evolve, open clusters undergo
tidal evaporation into the bulk of the stellar populations of the
Galactic disk. This process is helped (Pfalzner 2009) if the cluster contains
$\sim$1,000 stars or more. With those numbers, a significant
population of O stars is expected, whose stellar winds, together with
supernovae from early-type stars, will expel the remaining gas,
disturbing the clusters' dynamical equilibrium. Given this strong
trend between cluster size and age, searches are often biased against
old (i.e. more dispersed) systems, harder to separate from the stellar
background. Hence, open clusters are usually found to contain bright,
blue stars, indicating young ages.

In addition to age, metallicity is a property that can reveal the past
history of stellar populations in clusters, disks and other components
of the Galaxy.  For example, Twarog et al. (1997) used metallicity to
study the local structure and evolution of the disk, and Friel et
al. (2002), used this property to determine the metallicity
gradient of the disk.  This type of study is also illustrated more
recently by Fu et al. (2009), and Magrini et al. (2009).  The first
group examine the radial gradient of the metallicity of open clusters
along the Galactic disk.  From this they create a model whereby
infalling gas grows the disk inside-out. Their model accounts for
current and past metallicity radial trends.  The second group find
evidence for inflow that varies radially, being lower in the outer
regions of the disk.  This clearly establishes metallicity as a key
parameter in deducing the past star formation history.

There are about 2,000 known Galactic open clusters, but because they
are difficult to observe, one should expect many more: Piskunov et
al. (2006) estimate the total number of open clusters could be of
order $10^5$.  A large catalogue of open clusters was recently
compiled from the literature by Dias et al. (2002b), henceforth
referred to as DAML.  The catalogue includes information about
the kinematics of the clusters as well as estimates from the
literature of their ages and metallicities.

Several authors have used orbit calculations to study open clusters.  Piskunov et al. (2006) considered the open cluster population in the all-sky catalogue ASCC-2.5(I/280A, Kharchenko, 2001) which is based on Tycho2, Hipparcos, and other catalogues.  From this, they calculate the open cluster orbits and identified four open cluster complexes, the members of which share kinetic behaviour and age.  For example, the youngest complex (age $<$79 Myr) is identified in the Gould Belt.  

Frinchaboy \& Majewski (2008) have made a study of open cluster 3-d kinematics.  The first phase, reported in their paper, starts with creating a uniform set of kinematic parameters.  For this, they measured the radial velocity of over 3,000 stars in open clusters and concentrated on those with Tycho2 proper motion values.  This amounts to 66 open clusters, with membership having been determined for radial velocity, proper motion, and spatial distribution.  Applications with orbits are due to follow.

Finally, and of most relevance to our work, is the study of Wu et al. (2009).  They use the previous release (2.9) of DAML, dated 2008 April 13.   Their data analysis addresses distances, proper motion, and radial velocity.  They adopt the DAML distances, but use the proper motion data from Kharchenko's ASCC-2.5 catalogue (Kharchenko, 2001).  They find the effect of the re-reduction of Hipparcos data by van Leeuwen (2007) to be less than the error values in DAML.  Regarding radial velocity, they substitute more recent values of Frinchaboy \& Majewski (2008), and of Mermilliod et al. (2008) into their data set.  They use three different axisymmetric Galactic potentials  and calculate the orbits of over 400 open clusters for 5Gyr back in time.  The main conclusions related to orbits are that the velocity dispersion of open clusters increases with age, showing disk heating.  They also find that the orbit eccentricity and maximum height above the Galactic plane are affected most by uncertainties in distance.  Finally, they compare orbital eccentricity of globular clusters, disk giants, disk F and G dwarfs, and open clusters, and deduce that about 4\% of open clusters are thick disk objects.  

This work extends the above by concentrating on unusual clusters, rather than the bulk of the population.  In this paper we study the orbits of the DAML open clusters within the
Galactic potential.  Section 2 presents the available data on open
clusters, and describes the method of orbit integration.  In Section
3, we present the results of a catalogue data validation, based mainly
on orbit morphology, and self-consistency tests.  In Section 4, we
carry out an analysis of cluster orbits, and in Section 5 we
identify clusters that may not be of purely thin disk origin, but instead due to events such as the
impact on the Galactic disk of a high velocity cloud, of a globular
cluster, or a merger with a dwarf galaxy.

\section{Open clusters and orbit calculations}
\subsection{Catalogue data}
DAML was developed from an original source by Mermilliod (1995), now called WEBDA \footnote{http://www.univie.ac.at /webda/navigation.html} which is still being maintained and updated, and which, combined with SIMBAD's proper motion data, provides the necessary data for kinematic studies of open clusters.  These proper motion data are from Loktin \& Beshenov (2003), and Kharchenko et al. (2005).  WEBDA provides no errors on distances, but the SIMBAD proper motion data have errors attributed to them.

DAML is a compilation of the data in the literature, of which one advantage is a single file with all the key data on the open clusters, with the latest updated version 2.10\footnote{www.astro.iag.usp.br/$\sim$wilton}, released on 2009 February 17, which we use throughout this work.
The DAML catalogue contains entries for 1787 open clusters.  The data fields in the catalogue include position on the sky, distance, proper motion, radial velocity, metallicity, size, colour excess, Trumpler type, numbers of stars used in determining proper motion, metallicity and radial velocity, and where relevant, the error on these quantities.  The extent of kinematic information available is that, of these 1787 clusters, 1114 clusters have at least an entry for the distance to the Sun, and of these in turn, 481 have the data on distance, proper motion, and radial velocity, with 439 of these also having the age.  Information on metallicity is present for 111 of the set of 439.

 Of the set of 439, 86\% have an age below 0.5Gyr.  
The oldest reported in DAML, Berkeley17, is $\sim$ 10Gyr, the youngest are $\sim$1Myr.  The metallicity  [Fe/H] range is from $-$0.835 to 0.27 dex, with 20\% of values less than $-$0.2 dex.

Both WEBDA and DAML, through studies such as that of Kharchenko et al. (2005), rely on stellar observational proper motion data in surveys, such as Hipparcos (Perryman et al. 1997), Tycho2 (H$\o$g et al. 2000), UCAC2 (Zacharias et al., 2004), or 4M (Frinchaboy \& Majewski 2008).  Caution is needed with UCAC2 and 4M, because, according to Frinchaboy \& Majewski (2008), there may be systematic errors in the proper motion values.

Determination of errors in basic kinematic data is essential in assessing the strength of any conclusion resulting from orbit calculations.  Paunzen \& Netopil (2006) investigated the accuracy of other relevant open cluster data such as age and distance.  They used eight papers on data compilations, and WEBDA's  references, checking that no piece of data is used more than once.  This then represented data for 395 clusters, amounting to 6437 individual pieces of data.  They determined the distributions of the absolute errors in age and distance via a statistical analysis of the sample.  This shows the peak distribution for age and distance errors to lie in the respective intervals 40-60\%, and 5-10\%.  Paunzen \& Netopil (2006) also compared their result with the averages given in the 2005 October version of DAML, and conclude that if using the parameters from this catalogue, the expected errors are comparable to those obtained independently from the literature.  This provides support for using directly the uncertainty values from the catalogue, a view shared by L\'{e}pine et al. (2008).

We recognise that the open cluster membership in DAML is in no way volume, or age-limited, and no attempt is being made here to study the overall Galactic open cluster system.  We also recognise that the body of data on open clusters is changing, with new data being published between revisions by Dias and collaborators.  Some more recent high-quality data, for example on radial velocity measurements (e.g. Mermilliod, Mayor \& Udry, 2008, and Frinchaboy \& Majewski, 2008), have not yet been taken into account in the latest (2.10) revision of the catalogue.  This is an important area of improvement, as Bovy et al.(2009) point out that the impact of a single star's radial velocity on that of a moving group can be significant.  Recognising that incremental improvements to the data on open clusters are taking place, nevertheless, our work aims to reach conclusions on the set of hitherto observed Galactic open clusters for a clearly defined sample as provided in DAML, and we therefore do not include these isolated changes.

\subsection{Orbit calculation}

To calculate the (proper, rather than osculating) orbits of the clusters, we use a Galactic potential proposed by Fellhauer et al. (2006) for this type of calculation, and assume it to be fixed in time. This potential has also been used by several authors, for example Law et al. (2005), Fellhauer et al. (2007), Vande Putte \& Cropper (2009), and Vande Putte, Cropper \& Ferreras (2009) to track a variety of objects in the Galaxy.

Briefly, the potential is axisymmetric, and therefore uses cylindrical coordinates $R,\phi ,z$ where $R$ is the distance to the z 
axis, $\phi $ is the azimuthal angle between the x axis and the projection 
of the position vector on the xy plane, and $z$ the distance of the particle 
above the xy plane.

This potential has three parts, representing a bulge, halo, and disk:

\begin{equation}
\label{eq4}
\Phi \, = \,\Phi _{b} \, + \,\Phi _{h} \, + \,\Phi _{d} 
\end{equation}

The bulge potential is given by a Hernquist (1990) expression:

\begin{equation}
\label{eq5}
\Phi _{b} (r)\, = \, - \,{\frac{{GM_{b}} }{{r\, + \,a}}}
\end{equation}

\noindent
where $M_{b} \, = \,3.4\,\times 10^{10}\,M_{ \odot}  $, and $a\, = \,0.7$ 
kpc, and where $r$ is the Galactocentric radius ($r^2 = R^2 +
z^2$). A halo contributes to the potential in a logarithmic form:

\begin{equation}
\label{eq6}
\Phi _{h} \, = \,{\frac{{v_{0}^{2}} }{{2}}}\,\ln (R^{2}\, + \,z^{2}\,q^{ - 
2}\, + \,d^{2})
\end{equation}

\noindent
with $v_{0} \, = 186$km s$^{{\rm -} {\rm 1}}$, $q$ is a flattening parameter, 
taken as unity here, and $d\, = \,12$kpc. A Miyamoto-Nagai expression (1975) represents the 
disk contribution:
\bigskip
\begin{equation}
\label{eq3}
\Phi _{d} (R,z)\, = \, - \,{\frac{{GM_{d}} }{{{\left\{ {R^{2}\, + \,{\left[ 
{a\, + \,(z^{2} + b^{2})^{1 / 2}} \right]}^{2}} \right\}}^{1 / 2}}}}
\end{equation}

\noindent
where $M_{d} \, = \,10^{11}\,\,M_{ 
\odot}  $, $a = \,6.5$kpc, $b\, = \,0.26$kpc. For this potential, 
the distance from Sun to Galactic centre is set at 8 kpc, with a local circular 
velocity of 220km s$^{{\rm -} {\rm 1}}$.

All calculations are carried out with a fixed Cartesian Galactocentric 
right-handed coordinate system. The 
x-axis points in the direction of motion of the Local Standard of 
Rest (LSR).The y-axis points from the Galactic centre 
to the Sun, so that the z-axis points to the North Galactic Pole (NGP).  The current cluster positions (x, y, z) used in the calculations are obtained from the cluster 
heliocentric distance, and its celestial coordinates. The velocities of a cluster with respect to the LSR are determined from its proper motion and 
radial velocity, and from the Solar motion (U$_{\odot}$=10$\pm 
$0.4, V$_{\odot }$=5.2$\pm $0.6, W$_{\odot}$=7.2$\pm $0.4 km s$^{{\rm -} {\rm 1}}$ from Dehnen \& Binney (1998).   Within the LSR's own coordinate system, the U axis points towards the 
Galactic centre, the V axis points in the direction of motion of the LSR, 
and the W axis points to the NGP. The transformation from radial 
velocities and proper motions to motion in the LSR (U, V, W), uses the formalism of Johnson \& Soderblom (1987).  The elements of the transformation matrix of Johnson \& Soderblom have been updated to the International Celestial Reference System (ICRS), based on RA(NGP)= 192.85948$^\circ$, Dec(NGP)=+27.12825$^\circ$, and the ascending node of the Galactic plane on the equator,  $I_\Omega$= 32.93192$^\circ$ (ESA, 1997).  The 
transformation from velocities in the LSR to velocities in our 
fixed Galactocentric system ($dx /dt $$ =V+ $ 220 km s$^{{\rm - 
}{\rm 1}}$, $dy /dt$ = -U, $dz /dt$ = W) uses an LSR velocity of 220$\pm $15 km s$^{{\rm - 
}{\rm 1}}$. The origin of the LSR is 
placed at 8kpc from the Galactic centre (Binney \& Tremaine, 2008).

The relevant equations of motion are integrated using a fourth-order Runge-Kutta procedure implemented in a FORTRAN code, and the code tolerance parameters set so as to conserve total energy and angular momentum to better than 10$^{ - 10}$.

\begin{figure}
\includegraphics[bb= 50 50 554 770,angle=-90,width=84mm,clip]{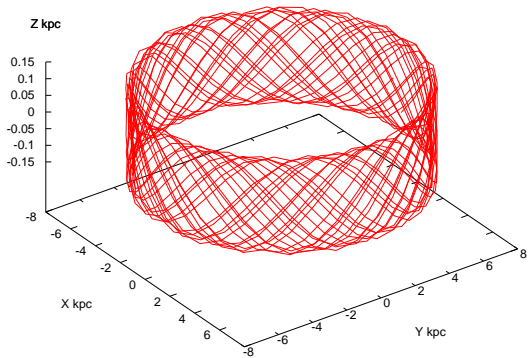}
\caption{Typical crown orbit, for open cluster NGC5316, calculated for 5Gyr only, for clarity}
\label{crown}
\end{figure}

We show in Table \ref{ext1} some examples of the observed data relevant to the orbit calculations, whereas Table \ref{ext2} displays the corresponding input data in the case of our Galactocentric representation.  In Table \ref{ext2}, the distance error is set at 10\%, and the error on the velocities also accounts for the error in the Sun's motion in the LSR, and in the circular velocity at the Solar radius.  All observed data in Table  \ref{ext1} are from DAML (version 2.10). \footnote{Since this work was submitted, new values for the Solar motion have been proposed (Sch\"onrich, Binney \& Dehnen, 2010): $(U,V,W)_\odot =
  (11.1_{-0.75}^{+0.69},12.24_{-0.47}^{+0.47},7.25_{-0.36}^{+0.37})$km s$^{-1}$,
  with additional systematic uncertainties $\sim(1,2,0.5)$ km s$^{-1}$.  We note that, $V_\odot$ is $7$ km s$^{-1}$ larger than previously estimated (see above), and this will have the greatest effect on $dx/ dt$ in Table \ref{ext2}.  However, given the magnitude of $dx/ dt$ and the size of the uncertainty in the velocity of the LSR ($\sim$15 km s$^{-1}$), little effect is found on the nature of the orbits, or on the orbital parameters z$_{max}$ and $\eta $ used in Table \ref{tabclus} (differences for the top seven candidates for external origin of a few percent of the uncertainties).}

\begin{table*}
\begin{minipage}{175mm}
\caption{Observed data relevant to orbit calculations for the first ten open clusters in DAML with sufficient data to calculate orbits, with data extracted from DAML.  A full version of the table for all 481 clusters for which an orbit calculation is possible, appears in the electronic version of the paper.}\label{ext1} \centering
\begin{tabular}{lcccccccc}
\hline Object & $\alpha (2000.0)$ & $\delta (2000.0)$ & $d_{\sun}$ &$v_{r}$& $\mu_{\alpha}\cos\delta$& $\mu_{\delta}$\\
 & h m s& \degr\, \arcmin\, \arcsec& (kpc)& (km s$^{-1})$& (mas yr$^{-1})$& (mas yr$^{-1}$)&\\ \hline
Berkeley 59&$00:02:14$&$+67:25:00$&$1.000\pm0.200$&$-12.5\pm
7.1$&$-2.11\pm0.81$&$-1.20\pm0.75$\\
Blanco 1   &$00:04:07$&$-29:50:00$&$0.269\pm0.054$&$  4.1\pm
1.4$&$20.17\pm0.51$&$ 3.00\pm0.51$\\
Alessi 20  &$00:09:23$&$+58:39:57$&$0.450\pm0.090$&$-11.5\pm 0.01$&$
8.73\pm0.53$&$-3.11\pm0.53$\\
ASCC 1     &$00:09:36$&$+62:40:48$&$4.000\pm0.800$&$-69.7\pm
4.7$&$-2.07\pm0.72$&$ 0.46\pm0.57$\\
Mayer 1    &$00:21:54$&$+61:45:00$&$1.429\pm0.286$&$-20.9\pm
2.0$&$-4.46\pm1.13$&$-6.66\pm0.94$\\
NGC 129    &$00:30:00$&$+60:13:06$&$1.625\pm0.325$&$-37.4\pm
0.5$&$-1.06\pm0.94$&$ 1.60\pm0.94$\\
ASCC 3     &$00:31:09$&$+55:16:48$&$1.700\pm0.340$&$-37.0\pm
0.0$&$-1.92\pm0.61$&$-1.25\pm0.59$\\
NGC 225    &$00:43:39$&$+61:46:30$&$0.657\pm0.131$&$-28.0\pm
0.0$&$-4.95\pm0.76$&$-0.50\pm0.76$\\
NGC 188
&$00:47:28$&$+85:15:18$&$2.047\pm0.409$&$-45.0\pm10.0$&$-1.48\pm1.25$&$-0.56\pm1
.24$\\
IC 1590    &$00:52:49$&$+56:37:42$&$2.940\pm0.588$&$-32.5\pm
6.4$&$-1.36\pm0.23$&$-1.34\pm0.83$\\
\ldots&\ldots&\ldots&\ldots&\ldots&\ldots&\ldots&\\
\ldots&\ldots&\ldots&\ldots&\ldots&\ldots&\ldots&\\
\hline
\end{tabular}
\end{minipage}
\end{table*}

\begin{table*}
\begin{minipage}{175mm}
\caption{Present positions and velocities of open clusters in Table \ref{ext1}, as calculated for use in the orbit code. A full version of the table for all 481 clusters for which an orbit calculation is possible, appears in the electronic version of the paper.}\label{ext2} \centering
\begin{tabular}{lccccccc}
\hline
Object & $x$ & $y$ & $z$ &  $dx / dt$ & $dy / dt$ & $dz / dt$\\
         &\multicolumn{3}{c}{ (kpc)}&\multicolumn{3}{c}{ (km s$^{-1}$)}\\
\hline Berkeley59&$0.878\pm0.176$&$8.471\pm0.094$&$
0.087\pm0.017$&$219.7\pm16$&$-25.3\pm
5.1$&$  2.4\pm3.7$\\

Blanco1&$0.013\pm0.003$&$7.952\pm0.010$&$-0.264\pm0.053$&$217.2\pm15$&$13.4\pm
4.9$&$ -1.7\pm1.7$\\

Alessi20  &$0.398\pm0.080$&$8.207\pm0.041$&$-0.030\pm0.006$&$
206.5\pm15$&$-0.2\pm
3.2$&$ -1.6\pm2.3$\\

ASCC1     &$3.527\pm0.705$&$9.887\pm0.377$&$ 0.014\pm0.003$&$
181.4\pm17$&$-75.8\pm13.9$&$ 21.9\pm11.3$\\

Mayer1    &$1.244\pm0.249$&$8.702\pm0.140$&$-0.023\pm0.005$&$
223.8\pm16$&$-51.2\pm
9.1$&$-33.8\pm10.5$\\

NGC129    &$1.402\pm0.280$&$8.818\pm0.164$&$-0.072\pm0.014$&$
197.1\pm15$&$-34.7\pm
6.4$&$ 21.8\pm7.7$\\

ASCC3     &$1.459\pm0.292$&$8.843\pm0.169$&$-0.221\pm0.044$&$
200.6\pm15$&$-43.0\pm
5.2$&$  3.2\pm5.0$\\

NGC225    &$0.557\pm0.111$&$8.348\pm0.070$&$-0.012\pm0.002$&$
209.7\pm15$&$-38.0\pm
3.3$&$  6.6\pm2.4$\\

NGC188    &$1.590\pm0.318$&$9.027\pm0.205$&$ 0.780\pm0.156$&$
199.8\pm19$&$-43.6\pm11.8$&$-14.8\pm11.8$\\

IC1590    &$2.448\pm0.490$&$9.597\pm0.319$&$-0.320\pm0.064$&$
206.8\pm16$&$-44.6\pm
5.6$&$ -8.0\pm12.1$\\
\ldots&\ldots&\ldots&\ldots&\ldots&\ldots&\ldots\\
\ldots&\ldots&\ldots&\ldots&\ldots&\ldots&\ldots\\
\hline
\end{tabular}
\end{minipage}
\end{table*}

\section{Data validation}

In this section we present the results of an initial screening of the data relating to those clusters for which sufficient information is available for orbit calculations.

\begin{figure*}
\begin{center}
\begin{tabular}{c@{\hspace{1pc}}c}
\includegraphics[bb=60 60 554 760,width=6.5cm,scale=0.8,angle=-90]{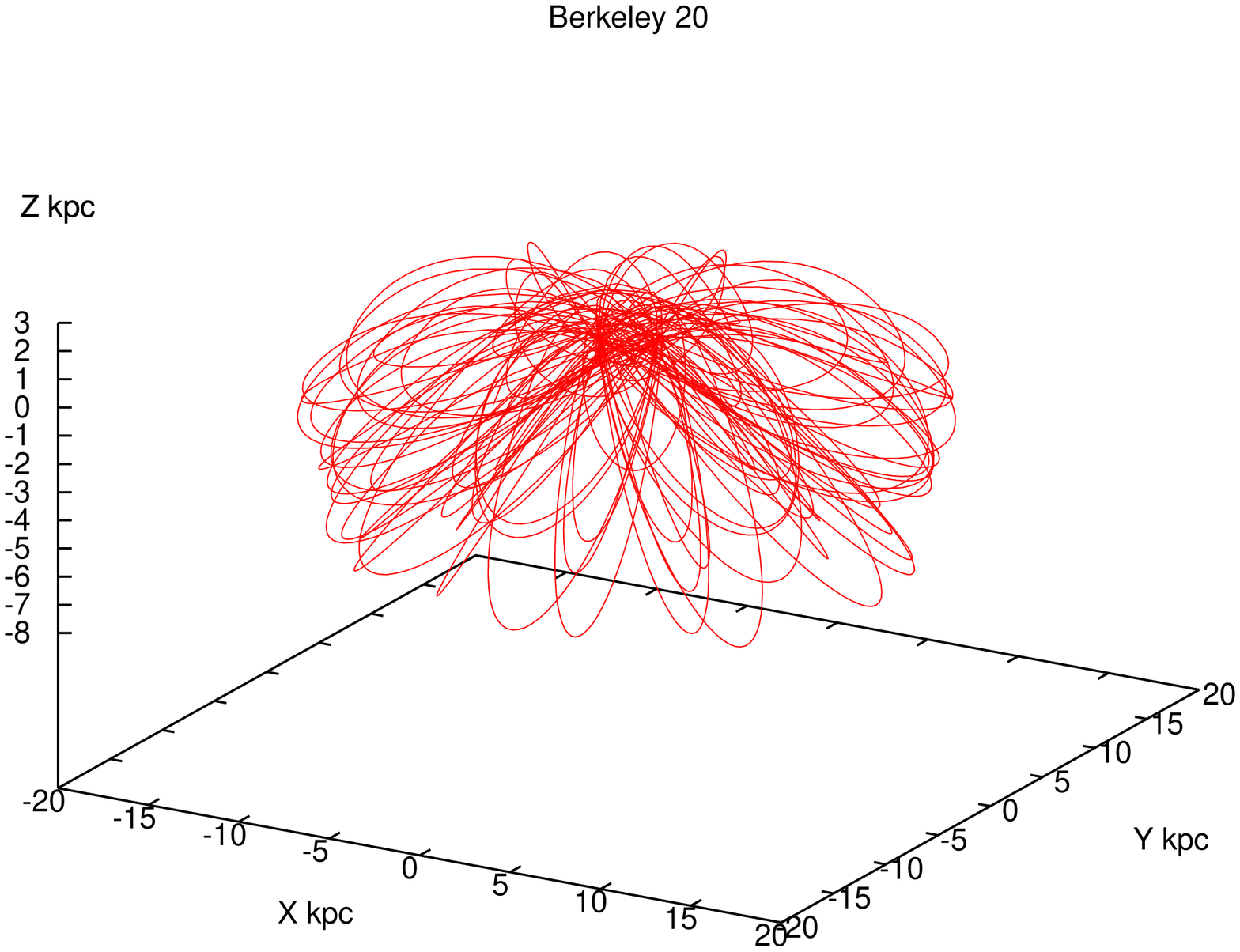} &
\includegraphics[bb=60 60 554 760,width=6.5cm,scale=0.8,angle=-90]{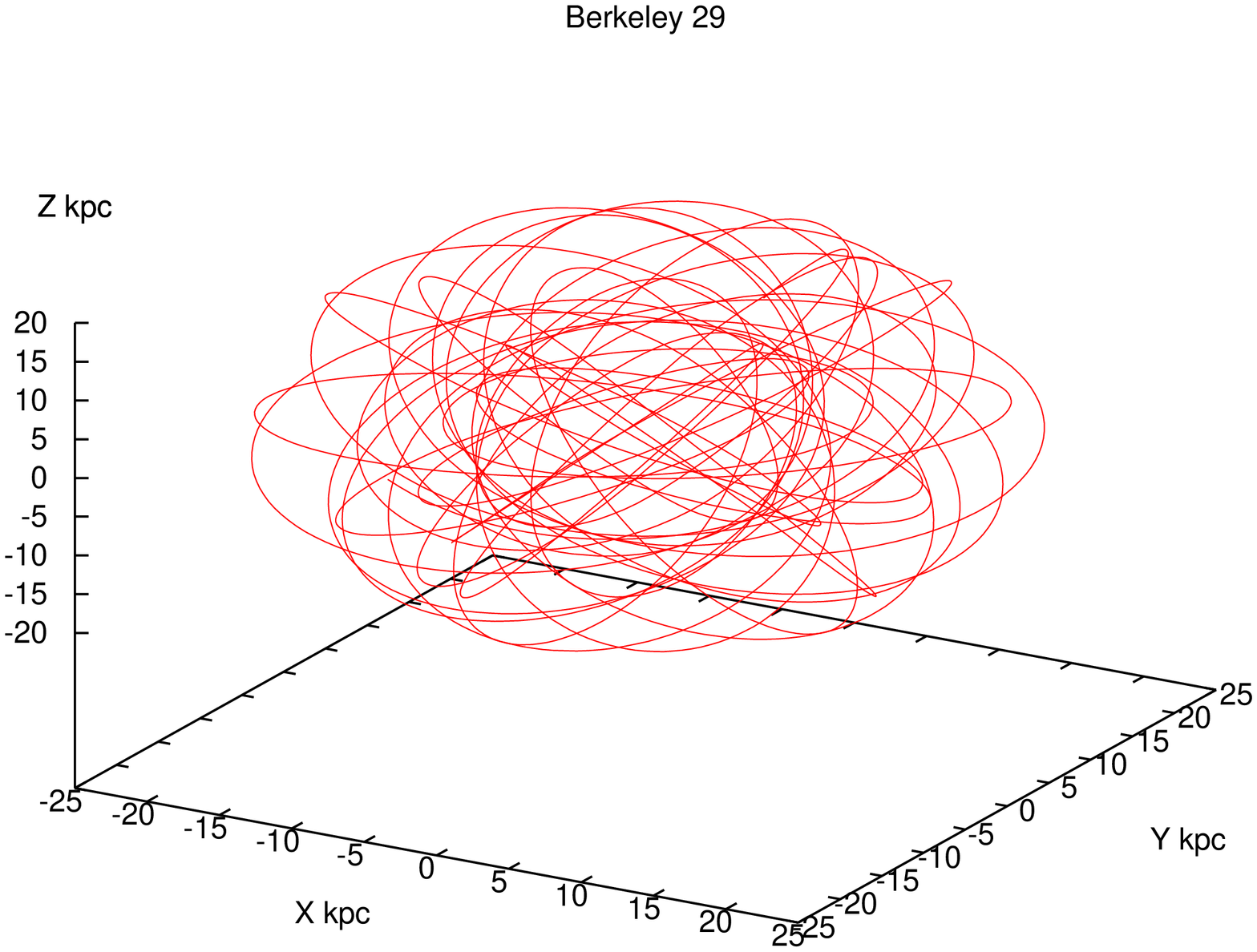} \\
\includegraphics[bb=60 60 554 760,width=6.5cm,scale=0.8,angle=-90]{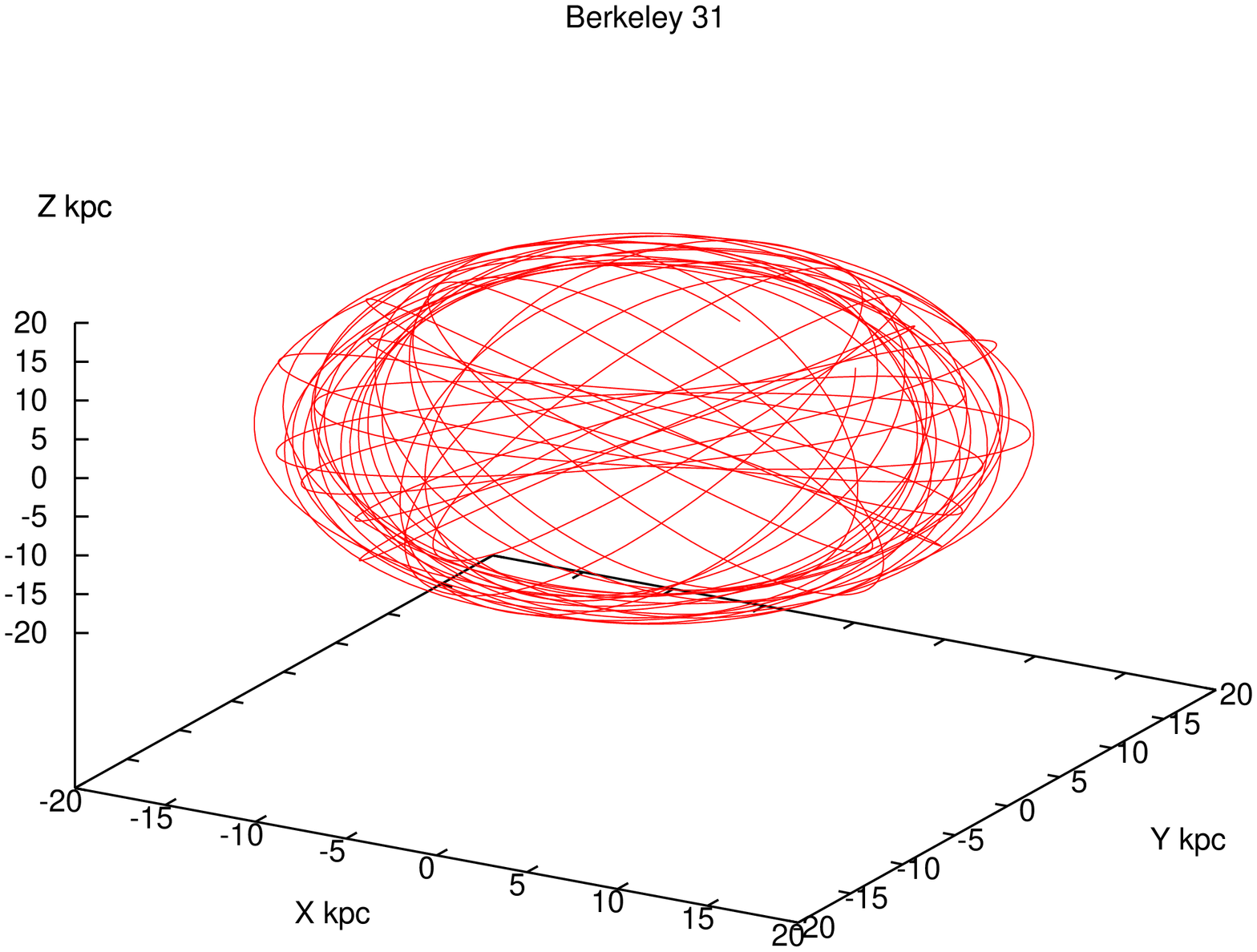} &
\includegraphics[bb=60 60 554 760,width=6.5cm,scale=0.8,angle=-90]{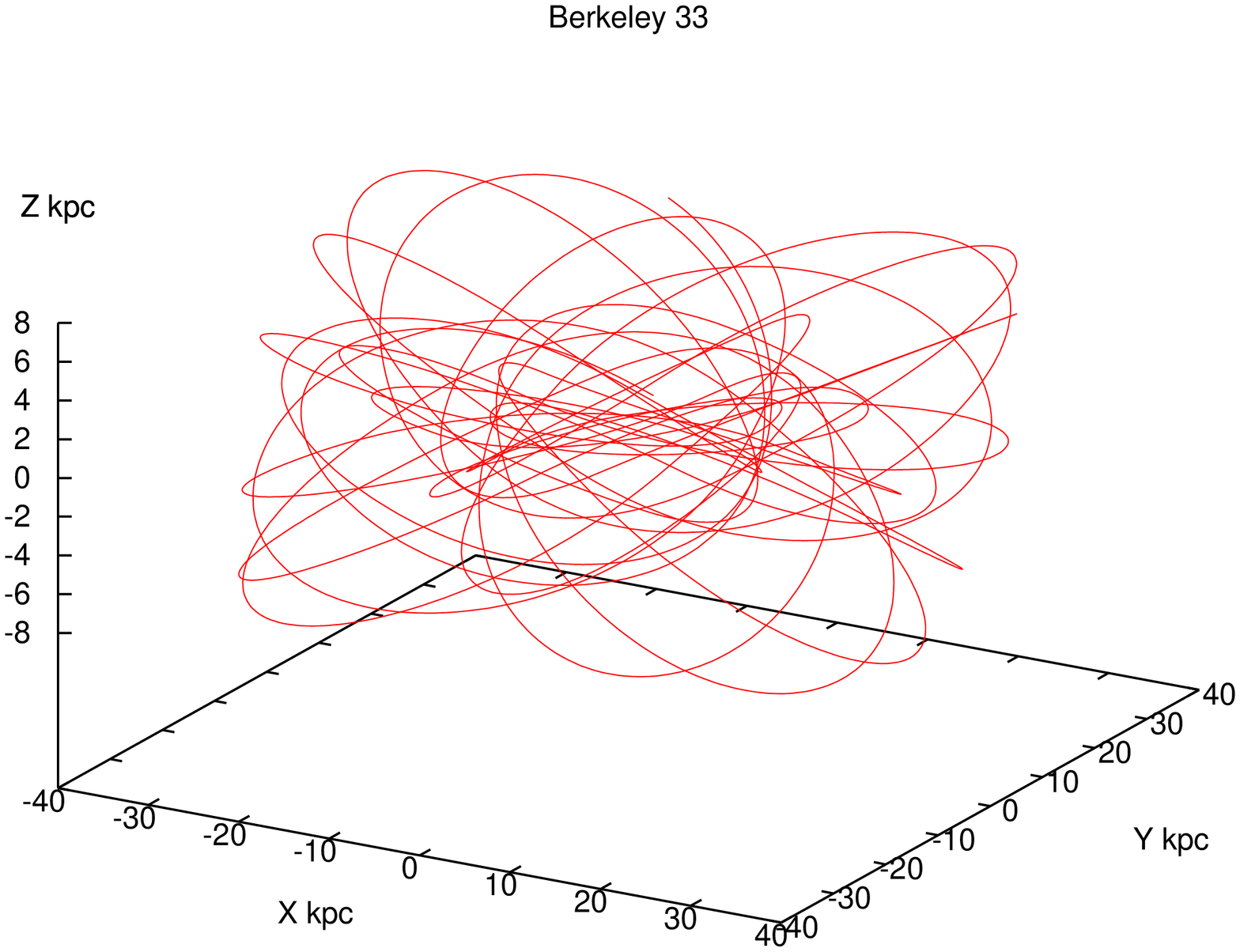} \\
\end{tabular}
\end{center}
\caption{Orbits for 15 Gyr back in time } 

\label{balls}
\end{figure*}

\begin{figure*}
\begin{center}
\begin{tabular}{c@{\hspace{1pc}}c}

\includegraphics[width=8cm,scale=0.7]{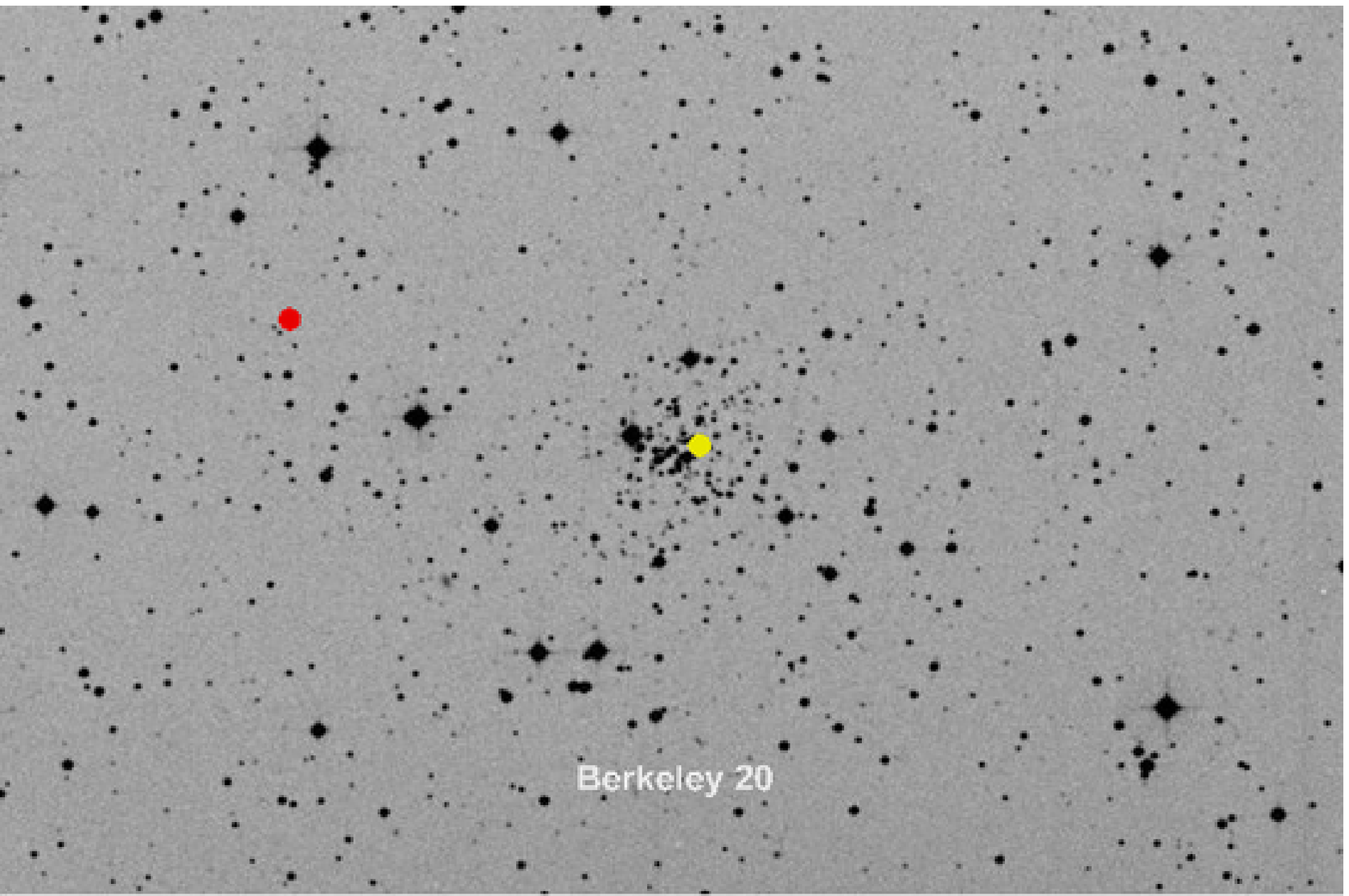} &
\includegraphics[width=8cm,scale=0.7]{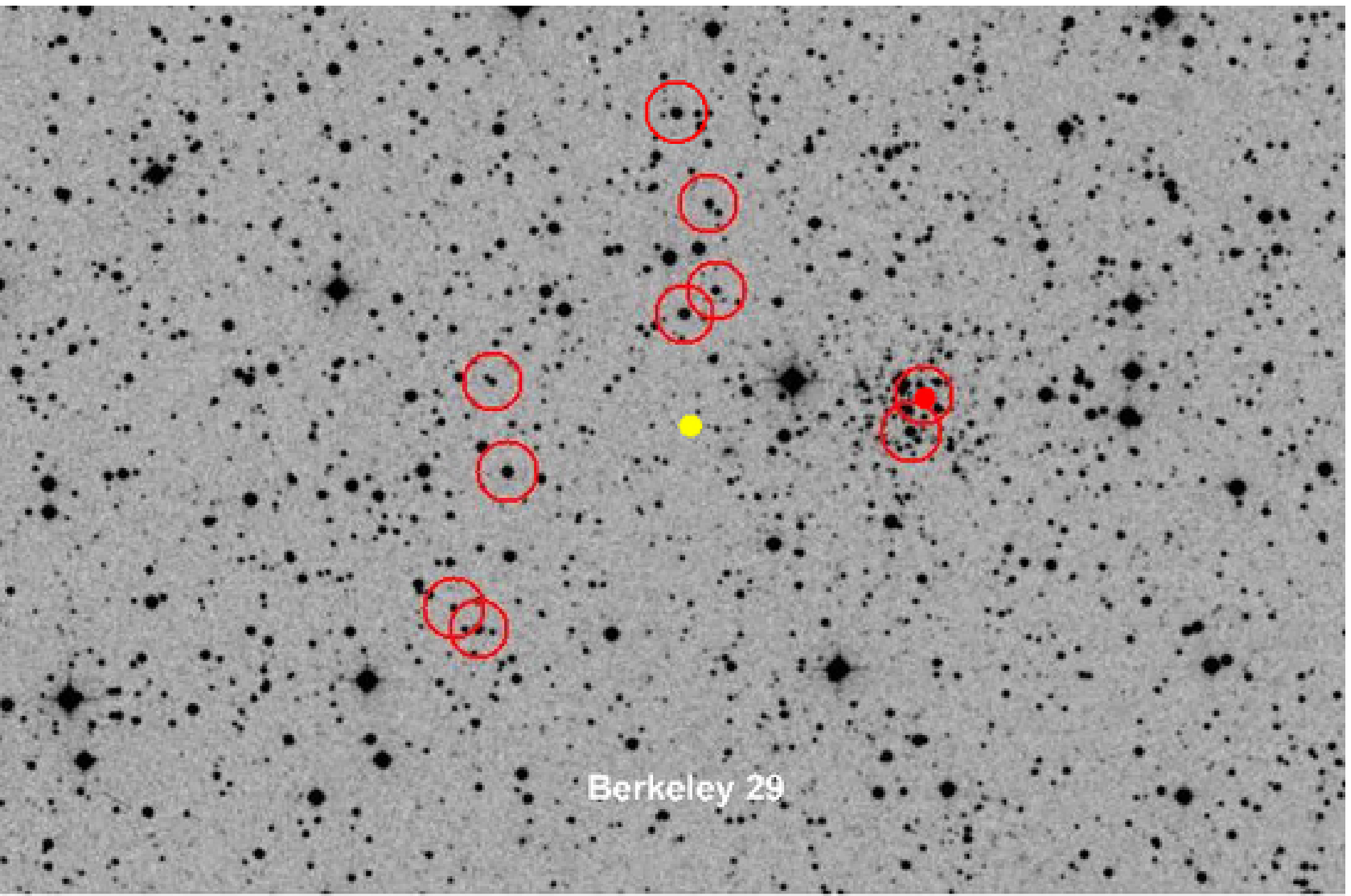} \\
\includegraphics[width=8cm,scale=0.7]{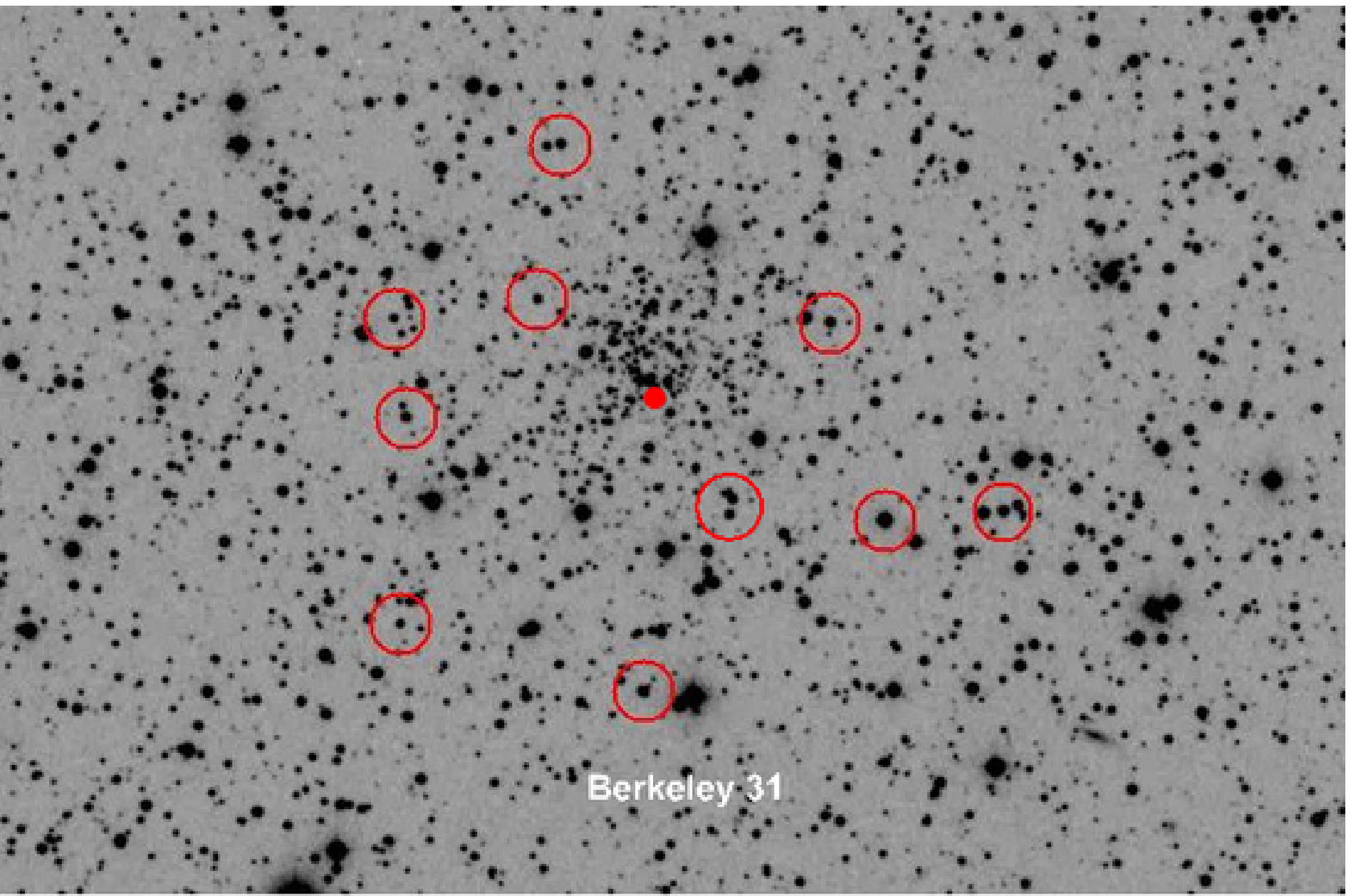} &
\includegraphics[width=8cm,scale=0.7,]{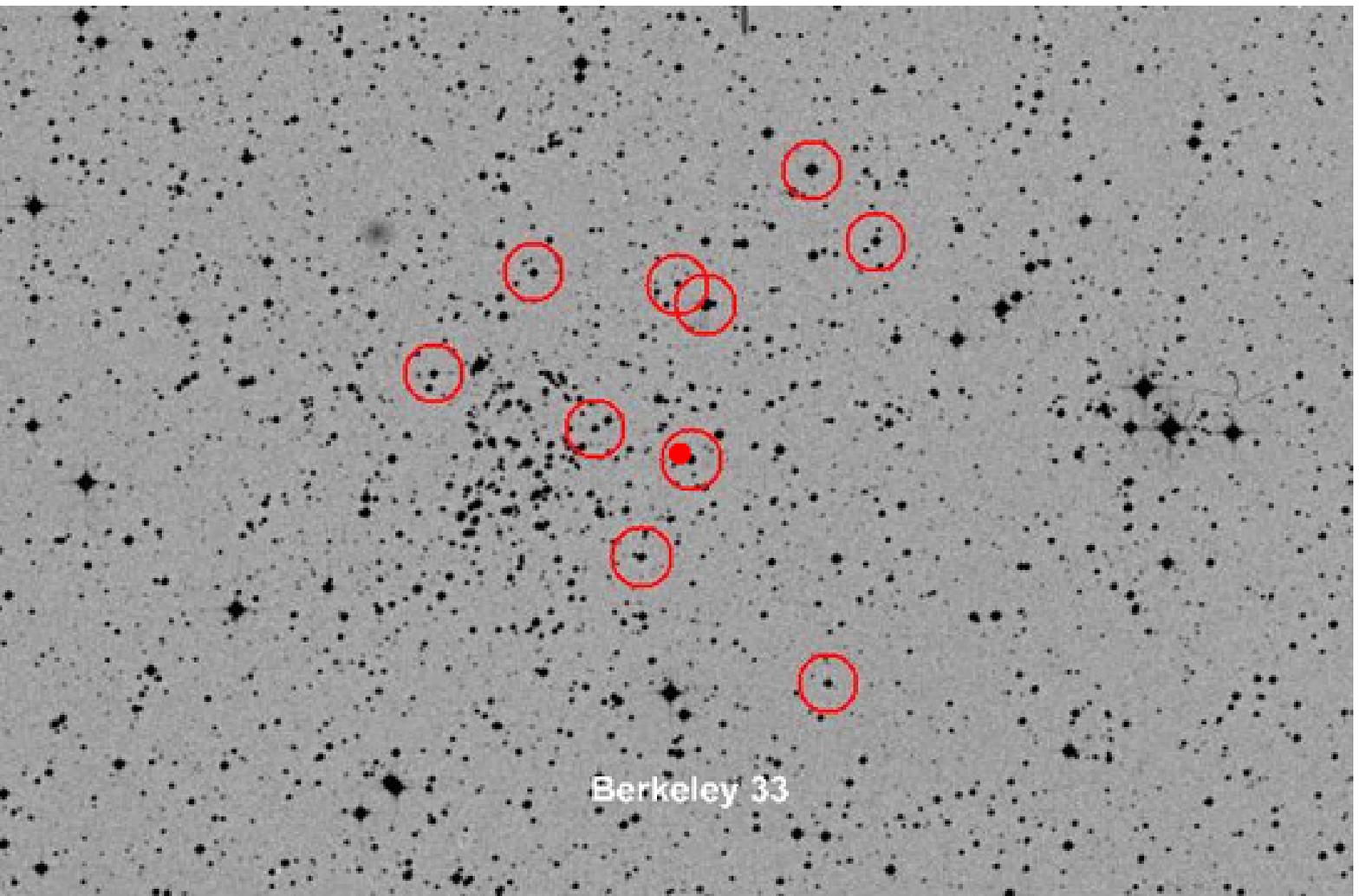} \\

\end{tabular}
\end{center}
\caption{Positions of the four clusters Berkeley20, 29, 31, 33 according to SIMBAD (red dot), and DAML (yellow dot).  SIMBAD and DAML positions coincide for both Berkeley31 and Berkeley33, so only a red dot is shown.  The red open circles have a radius of 25" and are centered on the 10 stars that have the highest probabilty of cluster membership.  There are no open circles for Berkeley20, as the proper motion stars lie $74^{\circ}$ off the nominal position of the cluster.  The figures are $\sim$19' wide and $\sim$13' high, and are sourced from the DSS at http:\/\/archive.stsci.edu/cgi-bin/dss\_form}. 
\label{find}
\end{figure*}

We considered the 481 clusters that  have at least distance, proper motion, and radial velocity data, and computed the orbits backwards for up to 15Gyr, visually examining each orbit.  The value of 15Gyr is longer than the age of the universe, and we recognise that the potential will have varied during this time, but this arbitrarily long time is used to place all cluster orbits on an equal basis.  Most clusters show quasi-periodic crown orbits, an example of which appears in Figure  \ref{crown}, in this case, for NGC5316.
There are four exceptions: Berkeley20, Berkeley29, Berkeley31, and Berkeley33 do not produce typical crown orbits.  Figure \ref{balls} displays their orbits after 15Gyr back in time and Figure \ref{find} shows sky images and positions of these four clusters. They also share the distinction of being the clusters reaching the highest vertical distance from the disk plane (called z$_{max}$), calculated over the orbit described in 15 Gyr (Table \ref{zmat}).  Furthermore, these clusters' absolute value of nominal $z$ at birth is highest of all, with z at birth being --3.8 kpc for Berkeley20, 13.9 kpc for Berkeley29, 12.9 kpc for Berkeley31, and 6.7kpc for Berkeley33.  The four clusters have remarkable properties: they are among the 10\% oldest for those 439 with known age, are among the seven most distant of our sample of 481, and have among the twenty lowest metallicity values in our sample of 111.  Clearly, the four clusters exhibit an interesting kinematic behaviour, meriting closer examination.  We summarise their properties in Table \ref{zmat}.

\subsection{Sense of Rotation}

We note that Berkeley29, one of the four clusters singled out in the pre-screening, counter-rotates with respect to the Galactic rotation: it is the only one of the 481 in the sample that does.  Uncertainties in the values of the Sun's motion in the LSR, 
velocity of the LSR, distances, radial velocities, and proper motions 
affect the cluster orbit (including its sense of rotation).  To examine the effect of these 
uncertainties on the sense of rotation of Berkeley29, we assumed that the 
uncertainties for Solar motion, LSR motion, distances, radial velocities and proper 
motions are all normally distributed, with standard deviations equal to the 
quoted experimental errors from the Dias database. We then performed a Monte Carlo simulation for Berkeley29, with 1,000 trials.  The result is that for 84\% of cases, the rotation is counter to that of the Galactic disk.  We note that Frinchaboy (2009) found a similar orbit to that of Figure \ref{balls}, also counter-rotating, using the kinematic information from the DAML catalogue.   While stars from a cluster originating in a cannibalised small companion could have counter-rotating orbit, counter rotation of stars is generally unusual, being observed in less than 1\% of disk galaxies (Sparke \& Gallagher, 2000).

\begin{table*}
\centering
\caption{The four clusters that do not describe crown orbits.  Columns 2 to 6 are from DAML}
\begin{tabular}
{|p{58pt}|p{53pt}|p{53pt}|p{53pt}|p{53pt}|p{53pt}|p{50pt}|}
\hline
Cluster& 
 $\mu _{\alpha} $ cos$\delta $& 
 $\mu _{\delta}  \quad _{}$& 
$d_{\sun}$& 
Age& 
[Fe/H] Solar& 
z$_{max}$ \\
%\hline
& 
(mas yr$^{ - 1})$& 
(mas yr$^{ - 1})$& 
(kpc)& 
(Gyr)& 
(dex)& 
(kpc) \\
\hline
Berkeley20& 
1.51$\pm $0.81& 
-4.11$\pm $0.82& 
8.4$\pm $1.68& 
6.02$\pm $1.51& 
-0.61$\pm $0.14& 
7.63$\pm $4.11 \\
%\hline
Berkeley29& 
-0.14$\pm $0.8& 
-4.75$\pm $0.58& 
14.87$\pm $2.97& 
1.06$\pm $0.26& 
-0.31$\pm $0.03& 
19.9$\pm $18.68 \\
%\hline
Berkeley31& 
-4.3$\pm $0.52& 
-3.97$\pm $0.52& 
8.27$\pm $1.65& 
2.05$\pm $0.51& 
-0.40$\pm $0.16& 
15.2$\pm $10.76 \\
%\hline
Berkeley33& 
-5.73$\pm $0.96& 
3.71$\pm $0.96& 
6$\pm $1.2& 
0.79$\pm $0.19& 
-0.26$\pm $0.05& 
7.97$\pm $11.21 \\
\hline
\end{tabular}
\\ \ Note that Carraro et al. (2007) give a very different value for the age of Berkeley29, namely 4.5Gyr.
\\
\label{zmat}
\end{table*}

\subsection{Coordinate Differences}

We notice that there is a small difference in coordinates between SIMBAD and the DAML catalogue for Berkeley20:  the SIMBAD coordinates (RA; Dec, J2000) are: $\alpha =05^{\rm h} 33^{\rm m} 00^{\rm s}$; $\delta = +00^\circ 
13\arcmin 00\arcsec$, whereas in DAML these values are $\alpha =05^{\rm h} 32^{\rm m} 37^{\rm s}$; $\delta = +00^\circ 
11\arcmin 18\arcsec$, which represents a separation of 5.99 arc min.  This, however, proves to be insufficient to produce the strange orbit in Figure \ref{balls}.

We notice that there is also a small difference in coordinates between SIMBAD and the DAML catalogue for Berkeley29:  the SIMBAD coordinates (RA; Dec, J2000) are: $\alpha =06^{\rm h} 53^{\rm m} 04\fs2$; $\delta = +16^\circ 
55\arcmin 39\arcsec$, whereas in DAML these values are $\alpha =06^{\rm h} 53^{\rm m} 18^{\rm s}$; $\delta = +16^\circ 
55\arcmin 00\arcsec$, which represents a separation of 3.35 arc min.  

For Berkeley31 and 33, the SIMBAD and DAML coordinates are identical.  Hence the cause of the odd orbits does not lie with the small coordinate differences.  This leaves the distance and proper motion as possible sources of error.

\subsection{Proper Motions}

The proper motion data in DAML are based mainly on the work of Dias et al. (2006), Kharchenko et al. (2003), and Loktin (2003), using the UCAC2 catalogue (Zacharias et al., 2004), the Tycho2 catalogue (H{\o}g et al, 2000), or the Hipparcos catalogue (Perryman et al, 1997), with cluster positions being identified by the centroid of a set of coherently-moving stars.  In the case of Berkeley20, 29, 31, 33, the proper motions are determined (respectively) by Dias et al. (2006) from UCAC2, by Loktin \& Beshenov (2003) from Tycho2 data, by Dias et al (2002a) from Tycho2, and by Dias et al. (2006) from UCAC2. 

Close scrutiny of the data relating to the individual stars used in DAML to determine the proper motion of Berkeley20 reveals that their position is $\sim 74^{\circ}$ away from the cluster.  This mismatch, probably due to a clerical error in the catalogue, explains the strange orbit.

\subsection{Distance}

\begin{figure}
\includegraphics[bb= 0 0 709 443,width=84mm,clip]{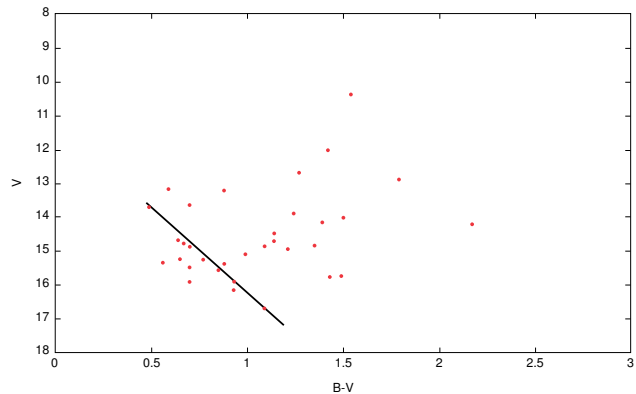}
\caption{CM diagram for stars used in proper motion determination of Berkeley29 in DAML.  The oblique runs through presumed cluster stars (see section 3.4)}
\label{dist1}
\end{figure}

There is some weak evidence for the distance to these particular clusters of ``proper motion'' stars being smaller than given in the catalogue for Berkeley29, 31, and 33.  Firstly, some estimate of the distance from Berkeley29 to the set of stars used for proper motion determination is possible using main sequence fitting.  Figure \ref{dist1} shows the B \& V magnitudes for these stars, based on the GSC2 catalogue (Lasker et al, 2008). Most stars are on the main sequence, and based on the position of the stars in the CM diagram, we believe that the stars along the oblique could be cluster stars, with the rest being field stars.  This oblique is almost parallel to the Hyades sequence, so this allows us to use the main sequence fitting method to find the distance to the cluster, based on the known distance to the Hyades cluster (see for example Karttunen et al., 1996).  This uses the relation\begin{equation}
\label{eq1}
V\, = \,M_{V} \, + \,5\log \left( {{\frac{{r}}{{10pc}}}} \right)\, + \,A_{V} 
\end{equation}where $V$ refers to the cluster at a given $B-V$, $M_V$ to the Hyades cluster at the same $B-V$, with Av determined from the NED database\footnote{http://nedwww.ipac.caltech.edu/} which incorporates the data of Schlegel et al. (1998).  We estimate the distance to the cluster containing the stars used in proper motion determination to be $\sim$1kpc rather than the DAML value of 15kpc (Table~\ref{zmat}).

Frinchaboy (2009) further concluded that the UCAC stars used to determine the proper motion of Berkeley29 are too bright to be part of that distant system.   A similar argument to that made for Berkeley 29 produces a distance of  $\sim$1kpc for the stars used in proper motion determination of Berkeley31.  No equivalent value is available for Berkeley 33, as no V magnitudes are given in GSC2 (Lasker et al, 2008) for the stars selected in the proper motion determination.

We then checked that reducing the distance to Berkeley29 to 1 kpc produces a crown orbit for the DAML proper motion and radial velocity, and we find that the cluster now rotates in the Galactic sense.  Likewise, reducing the distance to Berkeley31, and 33 to 1kpc also results in crown orbits.  This suggests that the proper motion values in DAML might refer to an association close by, distinct from Berkeley29 (or Berkeley31, or Berkeley33).

A second piece of evidence for a mismatch of distance and proper motion is provided by an estimate of proper motion for an object at the same distance and direction as the cluster, but in the disk, and assuming a constant circular velocity.  This is established using a formalism in Sparke and Gallagher (2000) and yields values of 0.01, 0.05, and 0.25 mas y$^{ - 1}$ for Berkeley29, 31, 33, which is 1--2 orders of magnitude less than the values given in the catalogue (see Table \ref{zmat}).

As a third piece of evidence, we note that Berkeley20, 29, 31, 33 are the most, second most, third most, and seventh most distant clusters from the Sun (Table~\ref{zmat}).  At these large distances, the cluster members may be too faint for the process of proper motion star selection to have detected them, and here we note that the limiting magnitude of UCAC2 (or even UCAC-3) is 16 (Zacharias et al., 2004, Zacharias et al., 2010, and Finch, Zacharias \& Wycoff, 2010).  This would then favour stars closer by. 

In principle, proper 
motion information for fainter stars are available in the 
USNO-B1.0 catalogue (Monet et al. 2003), and the recently released PPMXL catalogue (R\"{o}ser, Demleitner, Schilbach, 2010). However, their 
proper motion accuracy is not better than $\sim$ 4 mas 
yr$^{-1}$, which is larger than the expected proper motions 
for the cluster stars (see above). 

\subsection{Evidence from Non-kinematic Indicators}

The unusually old ages and low metallicities of Berkeley20, 29, 31, and 33 have been highlighted in the literature, for example in Yong et al., (2005), Carraro \& Bensby (2009), MacMinn et al. (1994), and Carraro et al. (2007). The values are included in Table~\ref{zmat}. 

\subsection{Adopted Sample}

Our orbit screening has  identified possible anomalies in the data for Berkeley20, 29, 31 and 33. Further analysis of the input data for the orbit analysis has shown that the data may be in error. We classify Berkeley20, 29, 31, 33 as outliers on-hold because of their orbit appearance, their high proper motion in relation to their distance in the Galactic potential, and because their large distance will bias  the process of selecting proper motion stars against the fainter clusters stars, in favour of closer associations of field stars.  For Berkeley29 and 31, proper motion stars could be foreground objects much closer to us. We plan to revisit these four clusters when newer data on proper motion have been carefully analysed and cluster proper motion measured.

Having found the small mismatch in coordinates between SIMBAD and DAML for Berkeley29, we then carried out a comparison between Simbad and DAML coordinates for the 368 out of 481 clusters in DAML clusters that are resolved by SIMBAD.  The comparison shows 3 cases where the difference between Simbad and DAML is in excess of 1$^{\circ}$: Markarian38 (113$^{\circ}$), Markarian6 (27$^{\circ}$), and ASCC14 (54$^{\circ}$).  We remove these three objects from further consideration, particularly as the SIMBAD database lists the Markarian objects as galaxies, and the third as a single, high-proper motion star.

Based on the observed large difference between the cluster position in DAML, and the centroid of the set of stars used to determine proper motion for Berkeley20($\sim 74^{\circ}$), we undertook a comparison of the celestial coordinates of the centroid of stars used to determine proper motion, versus the coordinates in the main DAML database.  To the extent that the positions of the stars used in proper motion studies are given in the Tycho2 (H{\o}g et al, 2000), UCAC2 (Zacharias et al, 2004), and Hipparcos (Perryman et al, 1997) catalogues, we were able to effect this comparison for 252 out of the 481 clusters.  We found that apart from Berkeley20 already noted above, there is only one other cluster where the difference exceeds 1$^{\circ}$: Bochum4 (130$^{\circ}$).  This cluster is also removed from further consideration.

Finally, we remove from consideration clusters for which less than five stars are used to determine proper motion.  This involves rejecting 63 clusters from our sample of objects with sufficient data to calculate orbits, leaving 410 clusters.  The resulting subset of clusters with also an age contains 369 members, of which 96 also have metallicity data.

\section{Orbit analysis}

We now look in more detail at orbits and metallicity in our cluster population.

Most clusters have orbits close to 
the Galactic plane. We expect such behaviour from clusters born from disk gas condensation, with their velocity lying in the plane of the disk, and aligned with 
the local circular velocity. 

A parameter of interest in this respect is z$_{max}$, the maximum distance of the orbit from the Galactic plane.  This shows whether the orbits could be associated with the thin or thick disk, or beyond.   
We can thus view  z$_{max}$ as a diagnostic of non-alignment with the Galactic plane velocities. 

We can also use the radial quantity

\begin{equation}
\eta \, = \,{\frac{{R_{max} \, - \,R_{\min}} } {{\left( {R_{max} \, + 
\,R_{\min}}   \right) / 2}}}
\end{equation}

where the variable {\it R} is the distance between the cluster and the Galactic 
centre, projected onto the disk. The parameter $\eta $  (twice the eccentricity for an elliptical orbit) will be small if the cluster velocity is aligned with the local circular velocity.  It is therefore also a diagnostic for non-alignment with the circular velocity.  

\begin{figure*}
\begin{center}
\begin{tabular}{c@{\hspace{1pc}}c}
\includegraphics[bb=0 0 840 593,width=90mm]{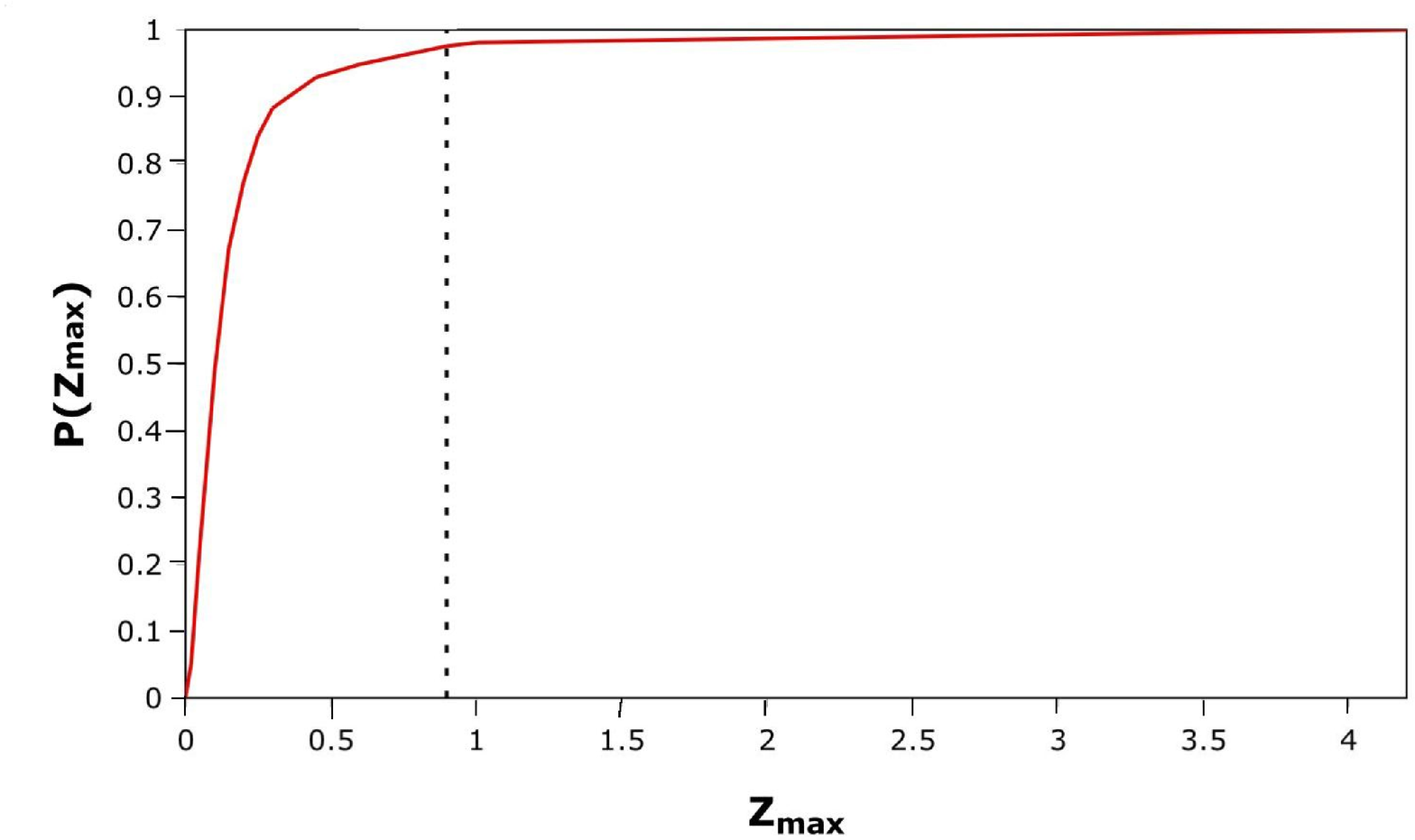} &
\includegraphics[bb=0 0 841 594,width=90mm]{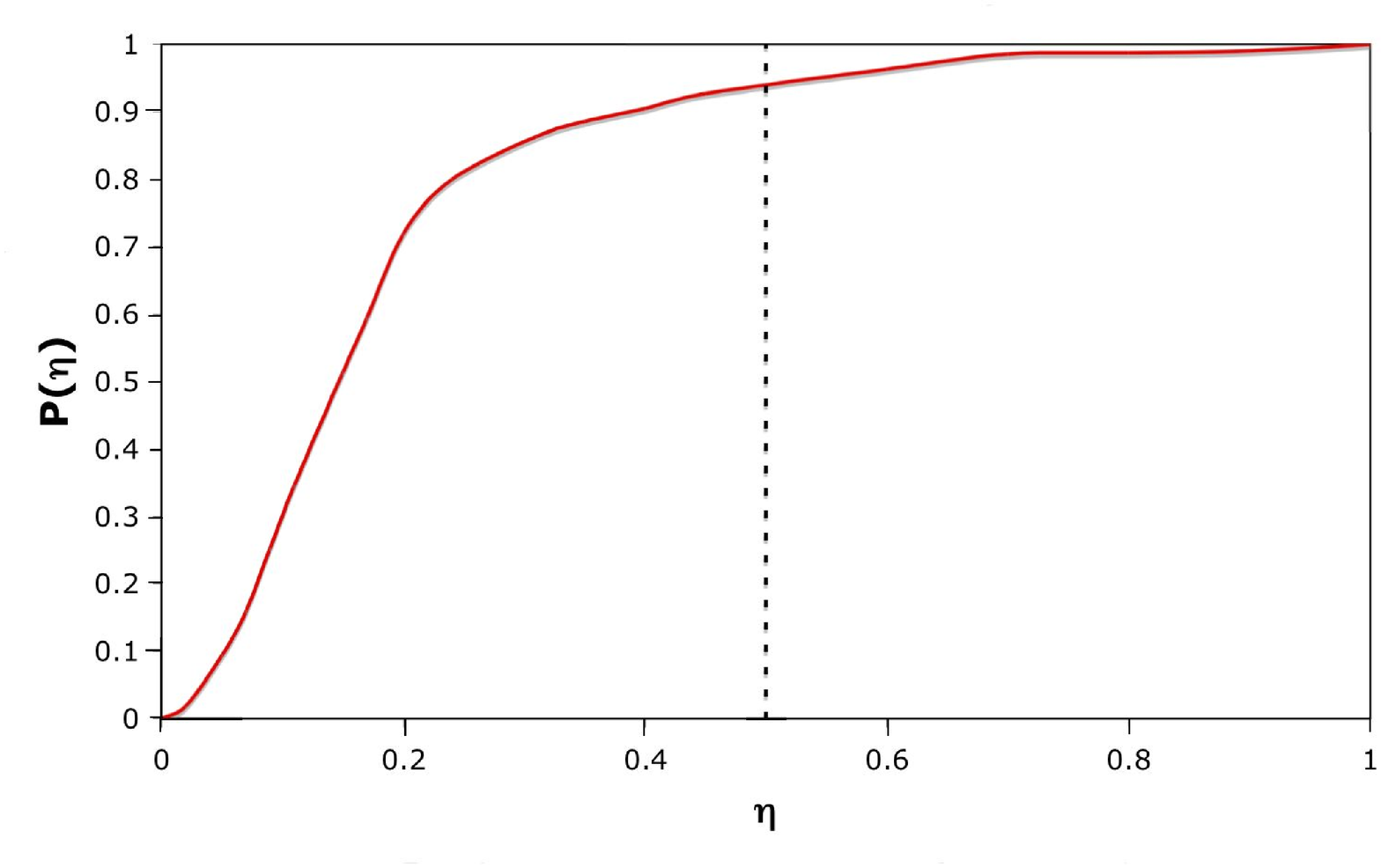} \\
\end{tabular}
\end{center}
\caption{Cumulative probability distribution of z$_{max}$, left panel, with threshold value of 0.9 shown by dashed line, and cumulative probability distribution of $\eta $ right panel, with threshold value of 0.5 represented by dashed line}. 
\label{cometz}
\end{figure*}

Figure \ref{cometz} shows the cumulative distribution of $\eta $ and z$_{max}$ values, illustrating the fact that most open clusters have low z$_{max}$ and $\eta $, as would be expected for an origin in the disk.  There are clear inflexions in the distribution, at z$_{max}$=0.35 kpc (90\% of clusters) and $\eta $=0.24 (80\% of clusters), indicating a possible transition to a different population.  In order to select a sample of the most extreme examples well beyond these inflexion points, we set the selection thresholds for segregating our results at z$_{max}$=0.9 and $\eta $=0.5.  These values correspond to z$_{max}$=3 times the scale height of the thin disk (Binney \& Merrifield, 1998), and $\eta $ when the orbit thickness in the disk plane is half the average distance between orbit and Galactic centre.  We consider that values above these thresholds will indicate an unusual origin.

The metallicity is clearly also a diagnostic of past history, and we also define a metallicity of --0.2 above which we find most clusters, to be a threshold below which we presume an unusual origin for a disk population.   

\begin{figure*}

\includegraphics[bb=0 0 710 442, scale=0.65]{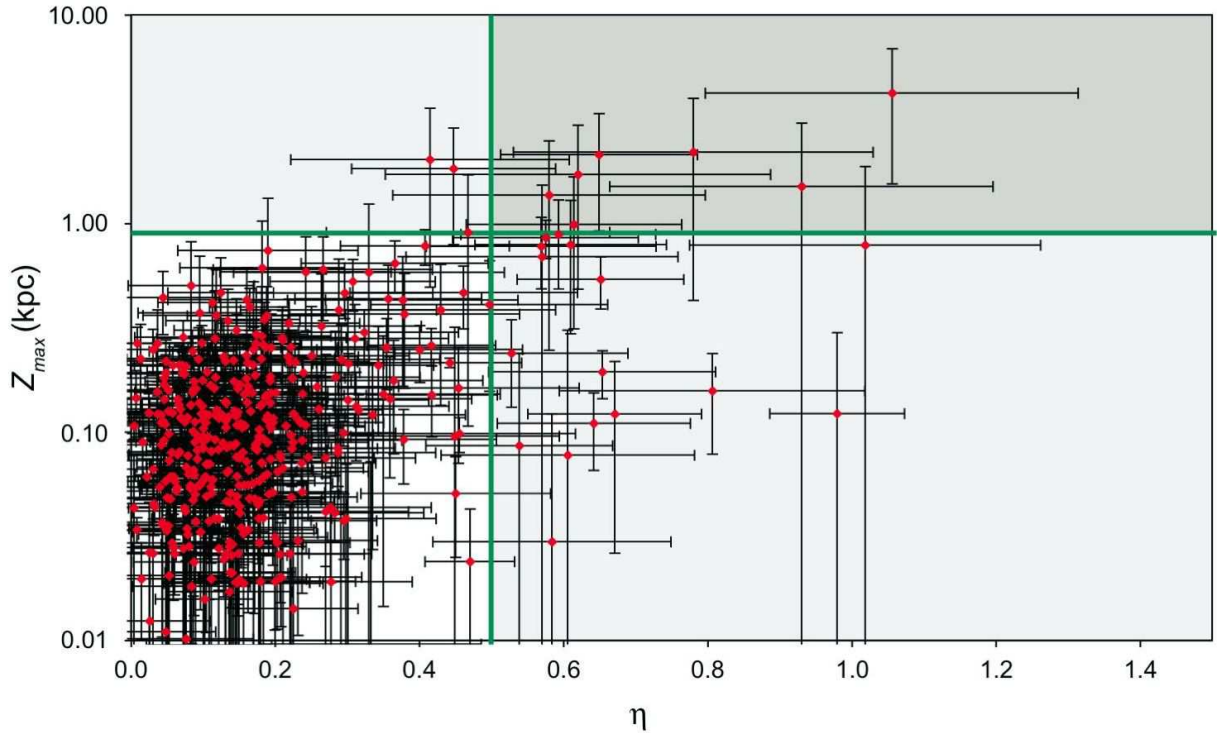}
\caption{Relation between z$_{max}$ and  $\eta $.  The green lines at z$_{max}$=0.9 and $\eta $=0.5 represent the thresholds beyond which unusual origin is a possibility.  The darker grey area represents the zone of foremost interest for clusters of unusual origin, with two criteria for unusual origin satisfied, the lighter grey corresponds to areas for which only one of the criteria is satisfied}
\label{zmaet10}
\end{figure*}

\begin{figure*}
\begin{center}

\includegraphics[bb=0 0 718 431,scale=0.5]{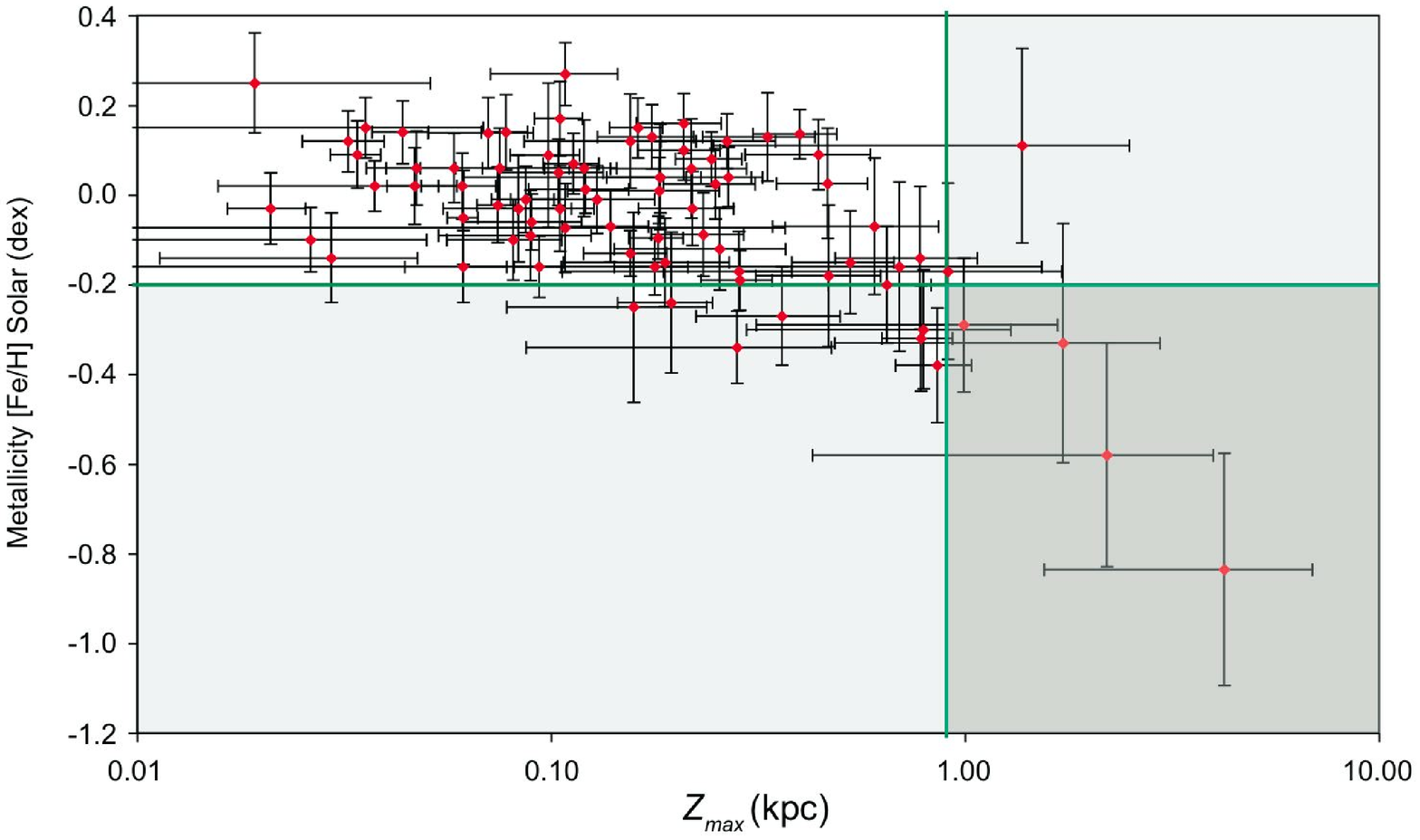}\\

\vskip0.2in
\includegraphics[bb=0 0 710 442,scale=0.5]{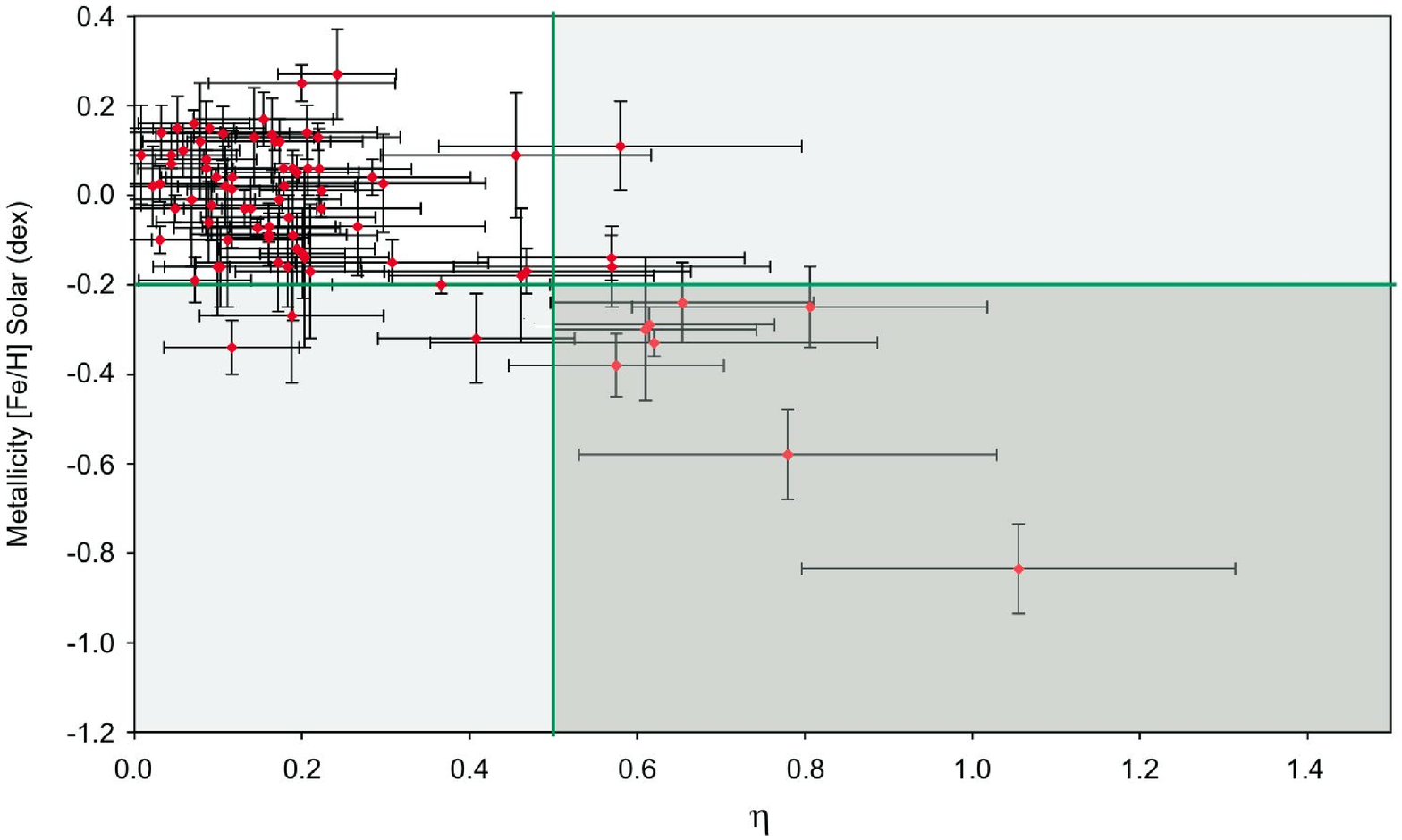} \\

\end{center}
\caption{Relation between z$_{max}$, and metallicity (top), and between  $\eta $ and metallicity (bottom)  The green lines represent the z$_{max}$ and $\eta $ thresholds beyond which unusual origin is a possibility.  The green line at metallicity $-$0.2 is the threshold below which unusual origin is a possibility.  The darker grey area represents the zone of foremost interest for clusters of unusual origin, with two criteria for unusual origin satisfied, the lighter grey corresponds to areas for which only one of the criteria is satisfied} 
\label{compclus}
\end{figure*}

We plot the various combinations of z$_{max}$,  $\eta $, and metallicity in Figures \ref{zmaet10} and \ref{compclus}.  Note that the metallicity error bars in Figures \ref{zmaet10} and \ref{compclus} represent the errors given in DAML, and in the case of  z$_{max}$ and $\eta $ are the standard deviation obtained from a Monte Carlo calculation carried out for 1000 realisations per cluster for the 410 clusters that satisfy the initial screening, using the error estimates in DAML as standard deviations of normal distributions, and a past orbit integration time of 15Gyr.  There is a correlation between all three parameters, z$_{max}$, $\eta $, and metallicity.

Table  \ref{tabclus} shows the main data for clusters that meet at least one of the criteria for unusual origin\footnote{As stated above, Wu et al., (2009) use an earlier version of the Dias catalogue complemented by some third-party proper motion data; they also use a different expression for the gravitational potential to the one used here (Allen \& Santill\'{a}n, 1991) so some differences in z$_{max}$ and $\eta $ can be expected, but we find these to be well within the uncertainties on our values.}. We see that of the 410 clusters initially considered, 35 satisfy at least one criterion.  Four satisfy all three criteria for unusual origin: Berkeley21, 32, 99, and Melotte66, and here the z$_{max}$ versus $\eta $ linear correlation coefficient is 0.97.  These clusters are aged 2.19, 3.38, 3.16, and 2.78 Gyr, respectively, which places them amongst the ten oldest clusters among the set of 369 for which ages are available in DAML.  Berkeley21 has the highest  z$_{max}$ and $\eta $, and its orbit for 15 Gyr is thick and wide. Thirteen others have data on metallicity, and of these, four meet two of the criteria, whereas nine meet only one criterion for unusual origin. We looked for information on cluster masses, but found only eleven clusters for which the mass is available (Durgapal \& Pandey, 2001), of which only two values are for our set of clusters in Table  \ref{tabclus}.  At least for these, the masses are not atypical in any degree.  In DAML, we find information on actual diameters in pc, but for the clusters in Table  \ref{tabclus}, we find no correlation between these diameters with age or metallicity.  

\begin{figure*}
\begin{center}
\includegraphics[bb=18 144 592 718,scale=0.7]{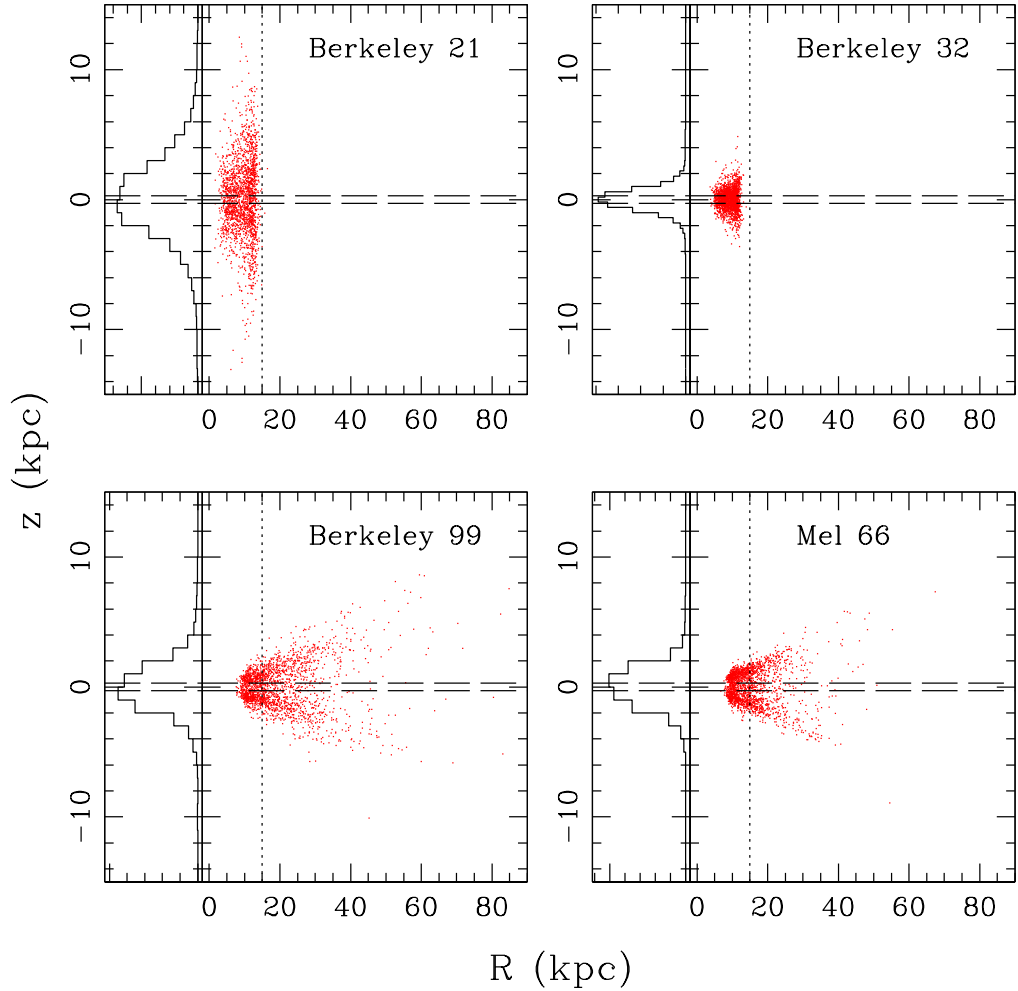} 
\end{center}
\caption{Monte Carlo realisations (2,000) for the birthplace of the four prime candidates for origin outside the disk.  The horizontal lines delimit the thin disk ($\pm$0.3 kpc), whereas the vertical line delimits the visible disk (15 kpc).}. 
\label{MC}
\end{figure*}

Finally, as support evidence, we have calculated the position of the birthplace of clusters which meet all three criteria using a Monte Carlo method with 2000 realisations.  Figure \ref{MC} shows the results in terms of $z$ and $R$ at birth for the four prime candidates for unusual origin.  For Berkeley21, there is a 92\% chance of birth occurring at $z\ge0.3$ kpc (one scale height).  For Berkeley32, this figure is 72\%.  For Berkeley99, there is a 90\% chance of birth beyond 0.3 kpc from the disk plane, increasing to 94\% if we also require that $R$ be less than the visible disk diameter of 15kpc.  For Melotte66, these figures are 88\% and 92\% respectively.   
The nominal values of $z$ at birth are for Berkeley21 2.5kpc, for Berkeley32  0.7kpc, for Berkeley99  --1.8kpc, and for Melotte66  --1.3kpc; these are among the ten highest distances from the plane calculated for those in our sample of clusters that also have age data.  In addition we carried out a similar calculation for NGC5316 (Figure \ref{crown}), and find that all values of $z$ at birth lie within the thin disk with a FWHM of 0.4kpc.

\section{Cluster Origin}

We now examine the nature of the unusual origin behind the clusters in Table \ref{tabclus} by considering the mechanisms that could lead to the formation of open clusters. We see that most clusters in Table  \ref{tabclus}, and certainly those where at least two criteria are met, exhibit relatively high values of z$_{max}$.  For high-altitude clusters such as these, two formation mechanisms have been suggested.  Martos, Allen, Franco \& Kurtz (1999) modelled the gas response to the spiral arm density wave and find that the shock sends gas from the inter-arm region to high altitude, followed by star formation, albeit with a low efficiency.  The other mechanism, proposed by de la Fuente Marcos \& de la Fuente Marcos (2008) has pre-existing clusters, preferably globular, interacting with high altitude gas clouds, creating enhanced turbulence leading to star formation in sufficiently large clouds.  They further estimate that about 3,000 clusters could have formed in this way over the life of the Galaxy. 

The above mechanisms could explain most clusters in Table \ref{tabclus}.  However, we offer further alternative mechanisms for consideration.  Firstly, as discussed by Friel (1995), impact of high velocity clouds on the disk can lead to star formation, a mechanism studied by Comer\'{o}n \& Torra (1992), with the cluster retaining some kinetic memory of the event, for example high $z$ (Danly, 1992) and possibly eccentric orbits.  In addition, Wakker et al. (1999) find that such High Velocity Clouds have low metallicities ($\sim$0.1 Solar), with some of them falling into the Galaxy.  These point to the origins of clusters with high z$_{max}$ and $\eta $ and low metallicity being formed as a result of an agent of extra-Galactic origin: in our case, these would be the top four clusters in Table \ref{tabclus}, and might also be the case for NGC2158, 2420, 7789, and IC1311.  Secondly, globular clusters impacting the disk can cause disk gas compression, either due to gravitational focussing (initially discussed in Wallin, Higdon \& Staveley-Smith 1996, and later in Vande Putte \& Cropper, 2009), or shock wave formation (Levy, 2000).  This would produce clusters with a metallicity representative of the disk, and the parameters z$_{max}$ and $\eta $ would be the indicators of unusual origin; in Table \ref{tabclus} these might be NGC6791, 1817, and 7044.  Next, the accretion of satellite galaxies could result in donations of ready-made open clusters with a metallicity higher than that of the high velocity clouds.  Here too, z$_{max}$ and $\eta $ would be diagnostics, because of the cluster retaining kinematic memory of the donor galaxy.  Metallicity could also be a diagnostic, but is likely to be higher than that of the high velocity clouds.  Finally, clusters could be formed as the result of a merging galaxy, such as within the dwarf Sagittarius or the Magellanic steams.  Here we would expect again high z$_{max}$ and $\eta $ and perhaps a higher or similar metallicity.  Some unspecified form of merger could be the agent responsible for the last nine clusters in Table \ref{tabclus}.  Clusters originating from these last two mechanisms should be related at least in terms of their integrals of motion to Galactic streams and/or to known HI features such as the Magellanic stream.

From the literature we find the following non-kinematic studies on the origins of these four clusters and for those we have placed ``on hold".  Concerning our four prime candidates for unusual origin identified by orbit analysis, we find that Yong et al. (2005) are unable to reach a conclusion for Berkeley21.  There are no further relevant studies for Berkeley32, 99, and Melotte66.  In the case of the clusters ``on hold",  Yong et al. (2005) using high resolution spectroscopy, provide arguments based on age and metallicity to conclude that Berkeley29 could be due to a merger, but not Berkeley31.  They are unable to reach a conclusion for Berkeley20 or 21.  Carraro \& Bensby (2009) reach the same conclusion in respect of Berkeley29, noting that it lies in the dSgr stream.  Carraro et al. (2007) conclude that Berkeley33 is not of extra-Galactic origin, but instead is a genuine Galactic cluster.  

While some small fraction of currently high Galactic altitude clusters could have been born at low altitudes, they would be expected to be of at least moderate metallicity.  For our four prime candidates this is not the case, supporting their origin high above the Galactic plane.

\begin{table*}
\centering
\caption{Main orbital parameters for clusters that meet at least one of the criteria for possible unusual origin.  A full version of the table (first 5 columns) for all 481 clusters for which an orbit calculation is possible, irrespective of how many criteria for unusual behaviour are met, appears in the electronic version of the paper.}
\begin{tabular}
{|p{65pt}|p{70pt}|p{50pt}|p{70pt}|p{70pt}|p{30pt}|p{30pt}|p{50pt}|}
\hline
Name& 
z$_{max}$& 
$\eta $& 
[Fe/H] Solar& 
Age& 
Stars& 
Crit.& 
Mech. \\
%\hline
& 
(kpc)& 
& 
(dex)& 
(Gyr)& 
& 
& 
 \\
\hline
Berkeley21& 
4.22$\pm $2.67& 
1.06$\pm $0.26& 
-0.835$\pm $0.1& 
2.188$\pm $0.547& 
32& 
z,e,m& 
Extra-Gal. \\
%\hline
Berkeley99& 
2.20$\pm $1.77& 
0.78$\pm $0.25& 
-0.58$\pm $0.10& 
3.162$\pm $0.791& 
5& 
z,e,m& 
Extra-Gal. \\
%\hline
Melotte66& 
1.72$\pm $1.24& 
0.62$\pm $0.27& 
-0.33$\pm $0.03& 
2.786$\pm $0.697& 
10& 
z,e,m& 
Extra-Gal. \\
%\hline
Berkeley32& 
0.99$\pm $0.68& 
0.61$\pm $0.15& 
-0.29$\pm $0.04& 
3.388$\pm $0.847& 
15& 
z,e,m& 
Extra-Gal. \\
%\hline
NGC2354& 
2.14$\pm $1.22& 
0.65$\pm $0.14& 
-& 
0.134$\pm $0.033& 
20& 
z,e& 
 \\
%\hline
NGC1893& 
1.51$\pm $1.51& 
0.93$\pm $0.27& 
-& 
0.003$\pm $0.001& 
15& 
z,e& 
 \\
%\hline
NGC6791& 
1.37$\pm $1.12& 
0.58$\pm $0.22& 
0.11$\pm $0.10& 
4.395$\pm $1.099& 
15& 
z,e& 
GC disk \\[2mm]
%\hline
NGC2158& 
0.16$\pm $0.08& 
0.81$\pm $0.21& 
-0.25$\pm $0.09& 
1.054$\pm $0.264& 
20& 
e,m & 
Extra-Gal. \\
%\hline
NGC7789& 
0.20$\pm $0.05& 
0.65$\pm $0.16& 
-0.24$\pm $0.09& 
1.413$\pm $0.353& 
10& 
e,m & 
Extra-Gal. \\
%\hline
IC1311& 
0.79$\pm $0.50& 
0.61$\pm $0.13& 
-0.3$\pm $0.16& 
1.585$\pm $0.396& 
25& 
e,m & 
Extra-Gal. \\
%\hline
NGC2420& 
0.86$\pm $0.18& 
0.58$\pm $0.13& 
-0.38$\pm $0.07& 
1.995$\pm $0.499& 
76& 
e,m & 
Extra-Gal. \\
%\hline
NGC2383& 
2.03$\pm $1.54& 
0.42$\pm $0.19& 
-& 
0.120$\pm $0.030& 
8& 
z& 
 \\
%\hline
Berkeley14& 
1.83$\pm $1.04& 
0.45$\pm $0.14& 
-& 
1.585$\pm $0.396& 
29& 
z& 
 \\
%\hline
NGC2324& 
0.91$\pm $0.80& 
0.47$\pm $0.20& 
-0.17$\pm $0.05& 
0.447$\pm $0.112& 
10& 
z& 
 \\
%\hline
Dolidze25& 
0.79$\pm $1.09& 
1.02$\pm $0.24& 
-& 
0.006$\pm $0.002& 
25& 
e& 
 \\
%\hline
ASCC63& 
0.12$\pm $0.18& 
0.98$\pm $0.09& 
-& 
0.017$\pm $0.004& 
10& 
e& 
 \\
%\hline
Pismis17& 
0.12$\pm $0.10& 
0.67$\pm $0.12& 
-& 
0.011$\pm $0.003& 
20& 
e& 
 \\
%\hline
Turner5& 
0.54$\pm $0.15& 
0.65$\pm $0.12& 
-& 
-& 
42& 
e& 
 \\
%\hline
ASCC43& 
0.11$\pm $0.04& 
0.64$\pm $0.13& 
-& 
0.191$\pm $0.048& 
19& 
e& 
 \\
%\hline
Trumpler16& 
0.08$\pm $0.23& 
0.61$\pm $0.18& 
-& 
0.005$\pm $0.001& 
8& 
e& 
 \\
%\hline
NGC436& 
0.89$\pm $0.40& 
0.59$\pm $0.13& 
-& 
0.084$\pm $0.021& 
17& 
e& 
 \\
%\hline
Bochum2& 
0.03$\pm $0.09& 
0.58$\pm $0.17& 
-& 
0.005$\pm $0.001& 
20& 
e& 
 \\
%\hline
NGC7044& 
0.69$\pm $0.84& 
0.57$\pm $0.19& 
-0.16$\pm $0.09& 
1.901$\pm $0.475& 
10& 
e& 
GC disk \\
%\hline
NGC1817& 
0.78$\pm $0.29& 
0.57$\pm $0.16& 
-0.14$\pm $0.05& 
0.409$\pm $0.102& 
10& 
e& 
GC disk \\
%\hline
Ruprecht47& 
0.09$\pm $0.15& 
0.54$\pm $0.13& 
-& 
0.078$\pm $0.019& 
7& 
e& 
 \\
%\hline
ASCC17& 
0.24$\pm $0.11& 
0.53$\pm $0.16& 
-& 
0.013$\pm $0.003& 
11& 
e& 
 \\
%\hline
NGC2506& 
0.65$\pm $0.18& 
0.37$\pm $0.13& 
-0.2$\pm $0.02& 
1.109$\pm $0.277& 
20& 
m& 
Merger \\
%\hline
King11& 
0.36$\pm $0.14& 
0.19$\pm $0.11& 
-0.27$\pm $0.15& 
1.117$\pm $0.279& 
10& 
m& 
Merger \\
%\hline
NGC869& 
0.18$\pm $0.06& 
0.21$\pm $0.11& 
-0.3& 
0.012$\pm $0.003& 
4& 
m& 
Merger \\
%\hline
NGC884& 
0.20$\pm $0.08& 
0.11$\pm $0.08& 
-0.3& 
0.013$\pm $0.003& 
18& 
m& 
Merger \\
%\hline
NGC6716& 
0.14$\pm $0.03& 
0.09$\pm $0.07& 
-0.31& 
0.091$\pm $0.023& 
11& 
m& 
Merger \\
%\hline
NGC2204& 
0.78$\pm $0.15& 
0.41$\pm $0.12& 
-0.32$\pm $0.1& 
0.787$\pm $0.197& 
19& 
m& 
Merger \\
%\hline
IC2581& 
0.28$\pm $0.19& 
0.12$\pm $0.08& 
-0.34$\pm $0.06& 
0.014$\pm $0.003& 
10& 
m& 
Merger \\
%\hline
NGC6451& 
0.43$\pm $0.18& 
0.16$\pm $0.11& 
-0.34& 
0.136$\pm $0.034& 
9& 
m& 
Merger \\
%\hline
NGC2451B& 
0.05$\pm $0.02& 
0.05$\pm $0.08& 
-0.45& 
0.044$\pm $0.011& 
5& 
m& 
Merger \\
\hline
\end{tabular}
\\ \S ``Stars" represents the number of cluster stars used to determine the proper motion.  ``Crit" is the criterion for unusual origin, with e representing $\eta$, z referring to z$_{max}$, m being metallicity.  A dash indicates that the information is not available.  ''Mech" refers to the formation mechanism, with Extra-Gal meaning an extra-Galactic agent has caused the cluster to form in the Galaxy, GC disk referring to a globular cluster impacting gas in the disk, and Merger referring to donation from a infalling galaxy.  The small horizontal gap in the table separates the clusters for which both $\eta$ and z$_{max}$ lie above the threshold for unusual origin.  Note that the A component in NGC2451 is a nearby moving group and the B component a background object.
\\
\label{tabclus}
\end{table*}

\section{Conclusion}

The DAML catalogue of 1787 open clusters provides a large database to study these clusters.  In particular, it contains sufficient data to calculate past orbits in as many as 481 cases.  This allowed us to undertake a systematic analysis of these open clusters by examining their orbit morphology.  

We found, with few exceptions, that the orbit type, when run for a sufficiently long time, is a crown orbit.  This type of orbit is readily characterised by two parameters: z$_{max}$,  the largest separation between cluster and disk in absolute value, and $\eta $, a measure of orbit eccentricity.  This has first of all revealed four clusters whose proper motion data in the catalogue are open to question, as they may relate instead to foreground associations: they don't exhibit a crown orbit, and one of them counter rotates.  These are Berkeley20, 29, 31, and 33, four otherwise interesting clusters with low metallicity and large distance from the Sun, but for which we found a mismatch between positional and kinematic information in DAML.  We temporarily place these ``on hold" for further study when more data on proper motion become available, as it is conceivable that our analysis has identified some confusion in the proper motion analysis reported in the catalogue.  This means that there may be four as-yet unidentified associations, generally closer to us than the four clusters listed above.

We then eliminate from the  cluster sample those with less than five stars (63 clusters) and four other clusters where data may be affected by clerical error.  We then examine the relation between z$_{max}$,  $\eta $, and metallicity in the resulting sample of 410.  

We suggest that the higher z$_{max}$ and $\eta $, and the lower the metallicity, the chances increase of the cluster being of unusual origin.  We fix thresholds of $>$0.9kpc, and$>$0.5 on z$_{max}$,  $\eta$.  If we also set a further threshold of $<$--0.2 in the metallicity [Fe/H] dex, we find four clusters breaching all three thresholds: Berkeley21, 32, 99, and Melotte66 which are the four strongest candidates for unusual origin: we argue this could be extra-Galactic (resulting from the impact of an external high velocity cloud).  An additional 24 clusters breach at least one of the thresholds, and we identify four other possible contenders for generation by HVC impact (NGC2158, 2420, 7789, and IC1311).  Three clusters may be due to globular cluster impact on the disk i.e. internal origin (NGC6791, 1817, and 7044).  Another four could be due to some form of merger.  These mechanisms are offered as alternatives to two previously suggested mechanisms for cluster formation at high altitude.

The error values given in the catalogue illustrate another important point.  Such uncertainties are bound to affect any study of the birth of open clusters, and is an example of how the ESA {\it Gaia} mission can help to substantially improve our understanding of our Galaxy's formation.

\section*{Acknowledgments}

We wish to thank the referee for numerous helpful and constructive comments that have helped improve the paper.  

This research has made use of the SIMBAD and Vizier databases, operated at CDS, Strasbourg, France.  This research has also made use of the NASA/IPAC Extragalactic Database (NED) which is operated by the Jet Propulsion Laboratory, California Institute of Technology, under contract with the National Aeronautics and Space Administration.  DVP acknowledges the support of an STFC grant.  TPG was funded during this work by a UCL MAPS faculty 2008 Dean's Summer Studentship.  RPM thanks STFC for support through a Rolling Grant.

\end{document}